\documentclass[preprint]{aastex}
\usepackage{natbib}
\newcommand{\etal}   {{\rm ~et al.}}

 \shortauthors{Kondratko\etal} \shorttitle{Accretion Disk
in NGC 3079}

\begin{document}

\title{Evidence for a Geometrically Thick Self-Gravitating Accretion Disk in NGC 3079}

\author{Paul T. Kondratko, Lincoln J. Greenhill, James M. Moran}

\affil{Harvard-Smithsonian Center for Astrophysics, 60 Garden St., Cambridge, MA
02138, USA}

\email{pkondrat@cfa.harvard.edu}


\begin{abstract}
We have mapped, for the first time, the full velocity extent of the water maser
emission in NGC 3079. The largely north-south distribution of emission, aligned
with a kpc-scale molecular disk, and the segregation of blue- and red-shifted
emission on the sky are suggestive of a nearly edge-on molecular disk on
pc-scales. Positions and line-of-sight velocities of blue- and red-shifted maser
emission are consistent with a central mass of $\sim2\times 10^6$\,$M_{\sun}$
enclosed within a radius of $\sim 0.4$\,pc. The corresponding mean mass density
of $10^{6.8}\,M_{\sun}$\,pc$^{-3}$ is suggestive of a central black hole, which
is consistent with the detection of hard X-ray excess ($20-100$\,keV) and an Fe
K$\alpha$ line from the nucleus. Because the rotation curve traced by the maser
emission is flat, the mass of the pc-scale disk is significant with respect to
the central mass. Since the velocity dispersion of the maser features does not
decrease with radius and constitutes a large fraction of the orbital velocity,
the disk is probably thick and flared. The rotation curve and the physical
conditions necessary to support maser emission imply a Toomre $Q$-parameter that
is $\ll 1$. Thus, the disk is most likely clumpy, and we argue that it is
probably forming stars. Overall, the accretion disk in NGC 3079 stands in
contrast to the compact, thin, warped, differentially rotating disk in the
archetypal maser galaxy NGC\,4258.

We have also mapped radio continuum emission in the vicinity of the disk and
identify a new, time-variable, non-thermal component (E) that is not collinear
with the previously imaged putative jet. Based on the large luminosity and the
unusually steep spectrum ($\alpha<-2.1$), we exclude a radio supernova as the
progenitor of E. However, because its spectrum is consistent with an aging
electron energy distribution, E might be a rapidly cooling remnant, which may
indicate that the jet axis wobbles. Alternatively, considering its location, the
component might mark a shock in a wide-angle outflow that is interacting with a
dense ambient medium. In this context, masers at high latitudes above the disk,
mapped in this and previous studies, may be tracing an inward extension of the
kpc-scale bipolar wide-angle outflow previously observed along the galactic minor
axis.
\end{abstract}

\section{Introduction}
NGC 3079 is a relatively nearby ($v_{sys}=1125 \pm 6$\,km\,s$^{-1}$), edge-on
spiral galaxy classified as Seyfert 2 \citep*[Ford et al.
1986;][]{Ho1997}\nocite{Ford1986} or a LINER \citep{Heckman1980}. Observations of
large-scale structure in this galaxy indicate the presence of a powerful outflow
that appears to originate within the nucleus. Along the minor axis of the galaxy,
at a position angle (P.A.) of $\sim80\degr$, a bipolar super-wind inflates a
kpc-scale superbubble visible in radio continuum emission \citep{Duric1988,
Baan1995, Irwin2003}, [NII]+$\mbox{H}\alpha$ emission \citep{Veilleux1994}, and
soft X-ray continuum emission \citep*{Cecil2002}. The apex of the superbubble is
embedded in a dense molecular kpc-scale disk traced by CO \citep{Koda2002}, HCN,
and HCO$^+$ emission \citep{Kohno2001}. The circumnuclear molecular disk is
aligned with, and rotates in the same sense as, the much larger galaxy disk such
that the north side is approaching and the near side is projected to the west of
its major axis.

There is compelling evidence that NGC 3079 contains an active galactic nucleus
(AGN). The X-ray spectrum exhibits a continuum excess in the $20-100$\,keV band
and a $6.4$\,keV Fe K$\alpha$ line, which are both believed to be unambiguous
indicators of nuclear activity \citep{Iyomoto2001, Cecil2002}. Both soft and hard
X-ray data are suggestive of unusually high hydrogen column density towards the
nucleus, $N_H\sim10^{25}$\,cm$^{-2}$ \citep{Iyomoto2001}, which, in the context
of the AGN unified model, is indicative of an almost edge-on obscuring structure
along the line of sight to the nucleus \citep{Lawrence1982, Antonucci1993}. The
existence of nuclear activity is also indirectly supported by radio observations
of the nucleus. Continuum images of the inner-pc region obtained with Very Long
Baseline Interferometry (VLBI) reveal a linear distribution of compact knots
consistent with a jet \citep{Irwin1988}. Furthermore, the galaxy harbors one of
the most luminous (i.e., assuming isotropic emission of radiation) water masers
known \citep{Henkel1984, Haschick1985}, and such maser emission generally is
associated with AGN activity \citep*{Braatz1997}. In fact, the isotropic
luminosity of the brightest, spatially unresolved spectral feature in NGC 3079
($\sim130\,L_{\odot}$) is comparable to luminosities of other AGN related masers
such as NGC 4258 and NGC 1068 but at least two orders of magnitude greater than
the isotropic luminosity corresponding to the total integrated maser flux from
starburst galaxies such as M82 \citep*{Baudry1994} and NGC 253 \citep{Ho1987} and
from the W49N maser, the most luminous maser in our galaxy associated with a
region of intense star formation \citep{Gwinn1994a}.

The nucleus of NGC 3079 may also host a starburst; the most convincing evidence
comes from infrared (IR) and far-infrared (FIR) observations. \cite{Lawrence1985}
found color excess and an extended $10\mu$m nuclear source indicative of large
amounts of dust in the nuclear region. The IRAS $60$ to $100\,\mu$m flux ratio is
$\sim0.5$, which is suggestive of warm dust on kpc scales and qualifies NGC 3079
as a starburst galaxy \citep{Soifer1989, Braine1997}. FIR observations of the
galaxy with ISOPHOT reveal a powerful $\sim3.0\times10^{10}\,L_{\sun}$ point
source of FWHM\,$\le4.5$ kpc coincident with the nucleus \citep{Klaas2002}. The
SED of this point source is indicative of $20-32$\,K dust temperature. The
luminosity of $3.5\times10^{10}\,L_{\sun}$ from IRAS \citep{Soifer1987} or
$\sim3.0\times10^{10}\,L_{\sun}$ from ISOPHOT \citep{Klaas2002} yields a star
formation rate of $\sim10$ $M_{\sun}$\,yr$^{-1}$ on kpc scales
\citep{Veilleux1994}, under the assumption that star formation alone contributes
to the infrared flux. The presence of nuclear star formation may also be inferred
from large quantities of molecular gas associated with the nucleus. The
observations of CO emission indicate a dynamical mass of
$7\times10^8$\,$M_{\sun}$ enclosed within a radius of $76$\,pc \citep{Koda2002},
as well as an unusually large molecular surface gas density relative to other
normal, starburst, or active galaxies of $\sim7200$\,$M_{\sun}$\,pc$^{-2}$
\citep[Planesas, Colina, \& Perez-Olea 1997;][]{Sakamoto1999,
Koda2002}\nocite{Planesas1997}. Although sufficient to fuel a starburst, such a
high concentration of gas does not necessarily imply intense nuclear star
formation \citep{Jogee2001}. However, the dynamical and gas masses deduced from
CO observations by \cite{Koda2002} are suggestive of a Toomre-$Q$ parameter that
is less than unity within 150\,pc of the nucleus and for isothermal sound speeds
$<50$\,km\,s$^{-1}$, consistent with the presence of molecular gas. Thus, the
unusually high concentration of gas in the nucleus of NGC 3079 is subject to
gravitational instabilities and might be actively forming stars. Detailed studies
of the kpc-scale superbubble are also indicative of the nuclear starburst. The
inferred starburst age ($>10^7$\,years) and gas depletion time
($\gtrsim10^9$\,years) are longer than the dynamical age of the superbubble
($\sim10^6$\,years), which is consistent with a starburst driving the bubble
expansion \citep{Veilleux1994}. Furthermore, based on (1) the similarity of the
superwind in NGC 3079 to that in prominent starburst galaxies like M82 and NGC
253 and (2) the general lack of correlation between the properties of outflow and
nuclear activity in a sample of galaxies, \cite{Strickland2004b} argue that
supernova feedback is responsible for the superbubble in NGC 3079. The
correlation between diffuse halo X-ray luminosity per unit stellar mass and the
infrared luminosity per unit mass in a sample of seven starburst and three
``normal" galaxies \citep[see Fig.4 in][]{Strickland2004b} is supportive of the
starburst hypothesis, although that correlation does not imply causation
particularly in the context of an alleged starburst-AGN connection.

Arguments have been made against the existence of a nuclear starburst in NGC
3079. The ratio of FIR luminosity to the molecular gas mass (i.e., the star
formation efficiency) computed on kpc scales is $L_{FIR}/M_{H_2}\approx13$, which
is greater than the value of $\sim4$ for the Milky Way and $1<L_{FIR}/M_{H_2}<7$
for Giant Molecular Clouds, but below the average value of $\sim 25$ observed for
starburst galaxies \citep{Young1986, Planesas1997}. In fact, small-beam
($6\arcsec$) $10\,\mu$m flux of NGC 3079 measured by Lawrence et al. (1985)
yields $L_{FIR}/M_{H_2}$ of only $4.7$ on smaller scales, which is suggestive of
low star formation efficiency in the nuclear region \citep{Hawarden1995}.
Furthermore, since the $10\,\mu$m flux in a $5.5\arcsec$ aperture is only 4\% of
the larger beam color corrected IRAS flux, the infrared emission is not centrally
concentrated, contrary to expectation for a nuclear starburst
\citep{Devereux1987}. The nuclear region of NGC 3079 ($< 500$\,pc) is
characterized by a flat spectrum between $10$ and $20$\,$\mu$m in contrast to
what is observed for starburst galaxies \citep{Lawrence1985, Hawarden1986,
Klaas2002}. According to \cite{Hawarden1995}, low extinction towards the parts of
the nucleus not obscured by the molecular disk, the lack of a $10$\,$\mu$m
compact source coincident with the relatively well-defined apex of the
superbubble, as well as the unusually high H$_2/$Br$\gamma$ ratio all argue
against the starburst hypothesis. However, the validity of these arguments may be
in doubt because of extremely heavy obscuration in the nuclear region due to dust
lanes and the massive kpc-scale molecular disk. Indeed, asymmetry in the K-band
image of the nucleus implies considerable extinction even at $2.2$\,$\mu$m
\citep{Israel1998}.

If the nuclear starburst hypothesis is accepted, then it is not unreasonable to
suppose the coexistence of star formation and nuclear activity in the inner
parsec of NGC 3079, although the particular relationship between the two remains
uncertain. It has been suggested that star formation and nuclear activity are
likely to coexist in galactic nuclei, either because they are coupled through
evolutionary mechanisms, or simply because they both depend on gas inflow and
accretion \citep[and references therein]{Fernandes2001}. The existence of star
formation and of young stellar populations with ages of $(7-20)\times
10^6$\,years in the circumnuclear tori of radii $\lesssim 1$\,pc has been
proposed to explain the UV excess of many Seyfert 2 galaxies
\citep{Terlevich1996}. Our galaxy contains a population of young, hot, bright,
and massive stars within the inner-pc of the central black hole (Genzel et al.
1996, 2003)\nocite{Genzel1996}\nocite{Genzel2003}, which might have formed
in-situ in a preexisting accretion disk \citep{Milosavljevic2004}. A young
stellar population with an age of $\sim7\times10^7$\,years has been found in the
central 12\,pc of the Circinus active galaxy \citep{Maiolino1998}, and young
massive stars with ages $\le10^7$\,years have been unambiguously detected within
$\sim 200$\,pc of $30\%$ to $50\%$ of Seyfert 2 nuclei surveyed \citep[and
references therein]{Mas-Hesse1994, Mas-Hesse1995, Fernandes2001}. Because of the
likely coexistence of star formation and nuclear activity in NGC\,3079, the
proximity of the galaxy, the presence of maser emission, and apparent jet
activity, NGC\,3079 is an important case to study with regard to understanding
connections between star formation and nuclear activity.

\subsection{Previous Imaging Studies of the Nucleus}
\subsubsection{Continuum}
VLBI continuum images of NGC 3079 (refer to Fig.\ref{comparison} for a schematic)
reveal several collinear components, A, B, C, and D, extending over $\sim
2.1$\,pc ($\sim 25$\,mas) at P.A.$\sim 125\degr$ \citep{Irwin1988, Trotter1998,
Sawada-Satoh2000, Sawada-Satoh2002}. A line connecting the continuum components
is significantly misaligned with respect to the superbubble axis but
approximately aligned with the southern edge of the bubble \citep{Cecil2001}.
Components A and B are separating at about $0.13\,c$, which strongly indicates
the presence of a jet \citep{Trotter1998, Sawada-Satoh2000, Sawada-Satoh2002},
while the dominant component B is stationary with respect to the molecular gas
supporting the maser emission to within a transverse speed of $0.02\,c$
\citep{Sawada-Satoh2000}. These observations led \cite{Sawada-Satoh2000} to
identify component B with the central engine. However, this interpretation is not
unique since, even if component B were a radio jet core, an offset of the core
from the central engine could be significant \citep[e.g., NGC
4258;][]{Herrnstein1997}. Component B would also remain stationary with respect
to the maser if B were shock excited emission, as would occur where a hot ionized
supersonic flow encounters a dense ambient medium \citep*[e.g.,][ Kellermann et
al. 2004]{Phillips1982, ODea1991, Aurass2002}. In support of the latter
hypothesis, \cite{Trotter1998} found that components A and B have similar
spectra, exhibit spectral turnovers between 5 and 22\,GHz, and consist of
multiple components or extensions. Since its spectrum is not flat, B is most
likely not a jet core. Instead, \cite{Trotter1998} suggested that both components
are lower redshift and lower luminosity counterparts of compact symmetric objects
whose flux densities have been observed to peak at gigahertz frequencies and
which are ascribed to the interaction between jets and associated high-density
nuclear gas.

The putative jet coexists with additional non-thermal sources of continuum
emission, whose nature remains ambiguous. Component E, identified by
\cite*{Kondratko2000}, and component F, identified by \cite{Middelberg2003}, both
lie $\sim1.5$ and $\sim3.7$\,pc approximately east of component B, respectively.
\cite{Middelberg2003} found that the separation of components E and B at 5 GHz
systematically decreases at a rate of $0.07\pm0.01\,c$ (i.e., roughly one beam
width over $\sim3$\,years). These authors suggested that E is stationary and that
component B is moving toward E. However, this proposition seems unlikely since B
appears to be stationary with respect to the extended molecular gas structure
underlying the maser emission. A more realistic scenario is that component E is a
pattern in a flow that only appears to move in the direction of B as its physical
characteristics change. Such apparently inward motions have been observed in
several quasar and AGN jets (Kellermann et al. 2004)\nocite{Kellermann2004}.

\subsubsection{Maser}
\begin{figure*}[!th]
    \centerline{\includegraphics[width=7in]{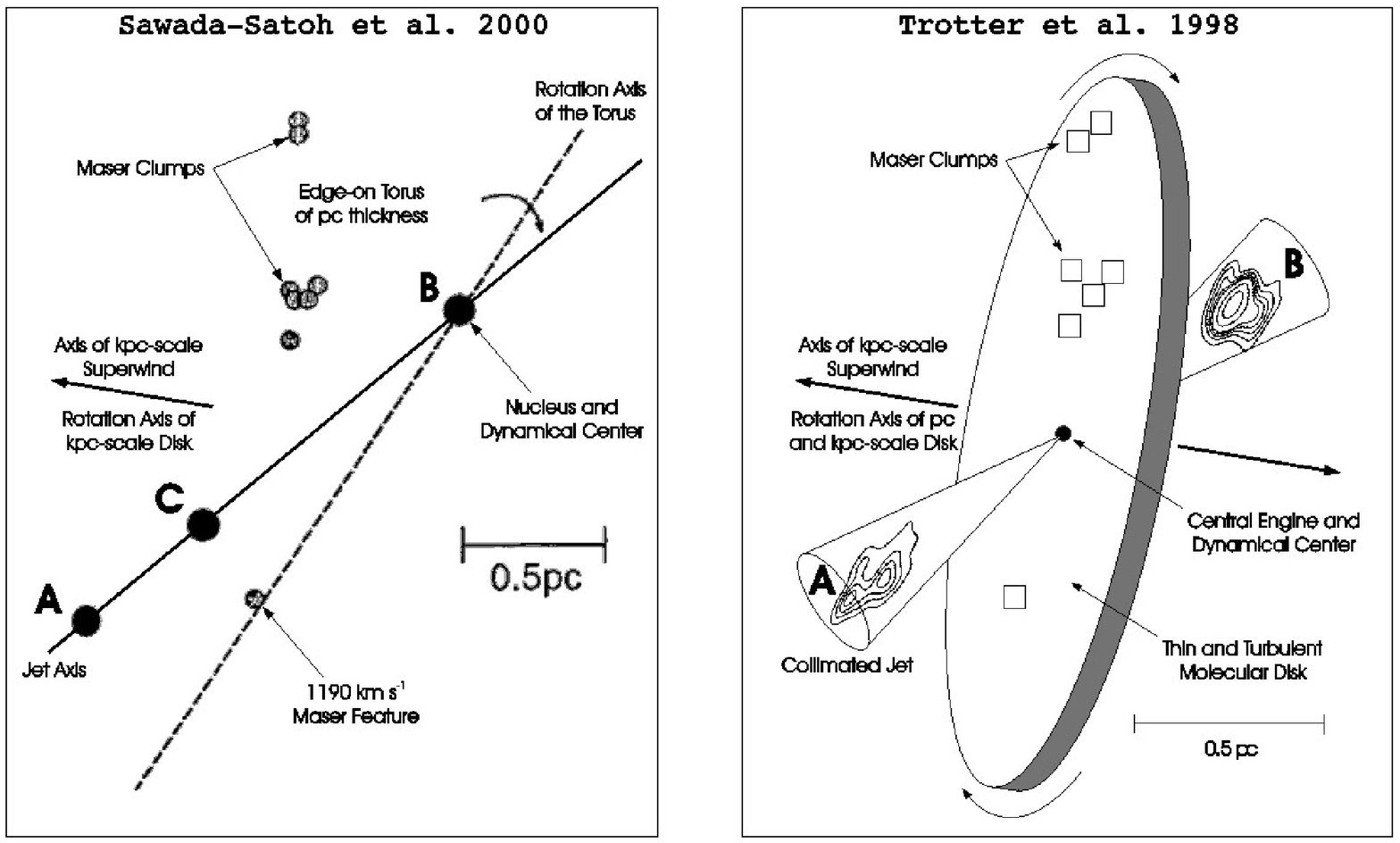}}
    \figcaption{Previously proposed models for the
    nuclear region of NGC 3079. In the model in the
    right panel,
    reproduced from \cite{Trotter1998}, the maser emission occurs in
    a thin, almost edge-on, and turbulent disk aligned with, and rotating in the same sense
    as,
    the kpc-scale molecular disk.
    In the model shown in the left panel, adapted from \cite{Sawada-Satoh2000},
    the maser emission takes place
    in a very thick, edge-on torus misaligned significantly with respect to the kpc-scale
    molecular disk. In both models, the outflow takes the form of a
    collimated jet not well aligned with the axis of the rotating structure.
    \label{comparison}}
    \hrulefill\
\end{figure*}

In previous observations (Fig. \ref{comparison}), the water maser in NGC 3079 was
found to be distributed in a disordered linear structure along
P.A.$\sim-10\degr$, aligned with the larger circumnuclear CO disk at
P.A.$\sim-11\degr$ \citep{Trotter1998, Koda2002}. Unlike the archetypal water
megamaser NGC 4258, the maser in NGC 3079 displays significant structure
orthogonal to the general elongation and large velocity dispersion
($\sim30$\,km\,s$^{-1}$) in relatively compact areas on the sky ($\sim0.1$\,pc).
Although no simple rotation law fits the data well, the velocity distribution of
the maser emission is consistent with a binding mass of $\sim 10^6$\,$M_{\sun}$
and indicates that the pc-scale accretion disk rotates in the same sense as the
kpc-scale molecular disk. The maser predominantly traces the approaching side of
the disk and thus reveals considerable asymmetry with the blue-shifted emission
significantly dominating the detected flux density. In particular, the maser
emission extends roughly from $890$ to $1190$\,km\,s$^{-1}$ with most emission
lying between $930$ and $1060$\,km\,s$^{-1}$. In addition to the main
distribution of the maser emission at P.A.$\sim-10\degr$, \cite{Trotter1998}
found two maser features that define a second axis at P.A.$\sim128\degr$. Since
one of these two maser features coincides on the sky with component C and the
other is in close proximity to component B, they may constitute a second
population of water masers in which amplification of the continuum components
might play a role.

The difficulty in interpreting the maser data has resulted in two very different
models for the nuclear region in NGC 3079 (Fig. \ref{comparison}). Relying on the
apparent alignment of the distribution of the maser emission with the kpc-scale
molecular structure, \cite{Trotter1998} proposed a largely north-south,
relatively thin, turbulent, and highly-inclined accretion disk for the geometry
of the nuclear region (Fig. \ref{comparison}). In this model, the dynamical
center is located between components A and B, while the maser emission occurs on
the surface of the disk in shocks stimulated by the interaction of a nuclear wind
with the disk. The asymmetry in blue- and red-shifted flux density is attributed
to free-free absorption in the ionized wind, related to the off-axis outflow as
traced by components A, B, and C.

The model proposed by \cite{Trotter1998} differs significantly from that proposed
by \cite{Sawada-Satoh2000}. According to these authors, the observed velocity
drift of the maser feature at $1190$\,km\,s$^{-1}$ is much larger than the
velocity drift of the other maser features and is due to centripetal
acceleration. In a rotating system, only the maser features nearest in projection
to the rotation axis are expected to show large velocity drifts. Consequently,
the rotation axis would have to pass close to the $1190$\,km\,s$^{-1}$ feature
and through component B, the proposed location of the dynamical center. In this
model (Fig. \ref{comparison}), the masers are embedded in a torus of parsec
thickness misaligned with respect to the kpc-scale molecular disk by $\sim
110\degr$. It is not clear why the embedded maser emission should be distributed
along the observed locus. However, the outflow axis is misaligned with respect to
the minor axis of the pc-scale rotating structure significantly more in the
\cite{Trotter1998} model than in the \cite{Sawada-Satoh2000} model.

\subsection{Motivation}
The \cite{Trotter1998} and \cite{Sawada-Satoh2000} models for the nuclear region
of NGC 3079 differ primarily in the location of the dynamical center as well as
in the thickness and orientation of the rotating structure. It is thus not clear
what is the correct geometric model for the inner parsec, what stimulates the
maser emission, and why it predominantly traces the regions of the molecular disk
approaching the observer. Ambiguity also remains about the connection of the
continuum components and the disk and the central engine, about the relationship,
if any, between the nuclear activity and the starburst, and about the connection
between the pc and kpc-scale outflows in this galaxy. Our study of NGC\,3079 is
motivated by these outstanding questions and by the disagreements between the two
existing models.

\begin{figure*}[!th]
        \centerline{\includegraphics[height=3in]{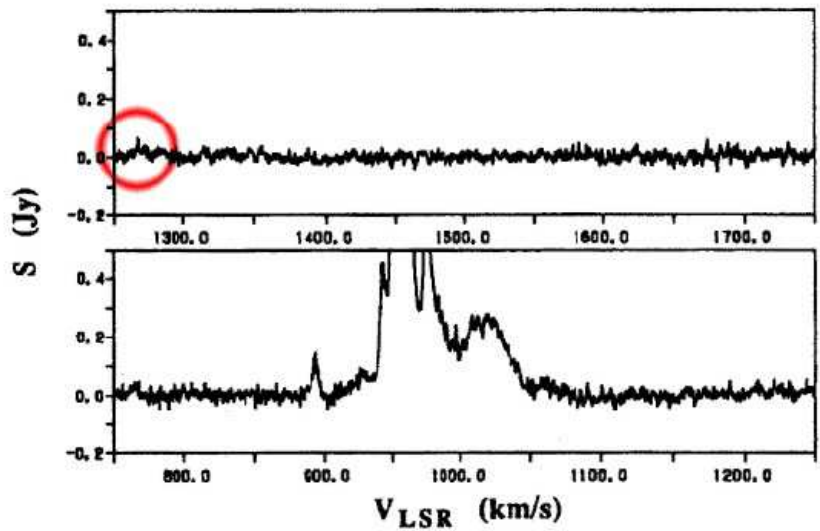}}
        \figcaption{First possible detection of red-shifted water maser emission
        in NGC 3079. Spectra reproduced from \cite{Nakai1995}.\label{Nakai}}
        \hrulefill\
\end{figure*}

In an effort to better understand the nuclear region in NGC 3079, we attempted to
fit a geometric and dynamical model of an inclined, thick, turbulent disk to the
maser positions and velocities reported by \cite{Trotter1998}. The model not only
reproduced the known emission at velocities $930-1060$\,km\,s$^{-1}$ but also
predicted the existence of emission at velocities $1190-1320$\,km\,s$^{-1}$
located roughly south from the known blue-shifted emission. No previous VLBI
observation had explored the velocities above $1222$\,km\,s$^{-1}$. However,
\cite{Nakai1995} marginally detected a $0.08$\,Jy peak at $1267$\,km\,s$^{-1}$
using the 45-m telescope of the Nobeyama Radio Observatory (Fig. \ref{Nakai}).
Motivated by this detection, we imaged with VLBI the water maser emission in NGC
3079 in order to determine whether the red-shifted emission lay south of the
blue-shifted emission as predicted by our model. Recently, more red-shifted
features have been detected with the Effelsberg 100-m antenna
\citep{Hagiwara2002}. In this study, we present the first VLBI map of the
red-shifted side of the accretion disk in NGC 3079, and thereby hope to clarify
the geometry of the inner-pc.

Data calibration and reduction techniques are discussed in Section
\ref{observation}. Spectral-line and continuum images of the inner-pc region are
presented in Section \ref{results}. In Section \ref{discussion} we interpret the
observed kinematics of the maser in terms of a thick, flared, disordered,
self-gravitating, and clumpy pc-scale disk, which orbits a central mass of
$\sim2\times 10^6$ $M_{\sun}$ enclosed within $0.4$\,pc. In this work we assume
that NGC 3079 is at a distance of $17.3\pm 1.5$\,Mpc \citep*{Tully1992}, so that
$1$\,mas corresponds to $0.084$\,pc. All velocities have been computed in
accordance with the radio definition of Doppler shift and in the local standard
of rest (LSR) frame.

\section{Observations and Calibration}
\label{observation} NGC 3079 was observed in spectral-line mode with the Very
Long Baseline Array (VLBA), augmented by the phased VLA and the Effelsberg (EB)
100-m antenna, for approximately 12\,hours on 2001 March 23. The source was
observed with $4\times 16$\,MHz intermediate-frequency (IF) bands, which covered
a velocity range of $726$ to $1491$\,km\,s$^{-1}$. The $5$, $8$, and $15$\,GHz
radio continuum observations were made on 1996 December 2 with the VLBA and the
VLA phased array. The source was observed with a $32$\,MHz bandwidth in each
polarization and for approximately three hours at each frequency. Both spectral
and continuum data were processed with the VLBA correlator at the
NRAO.\footnote{The National Radio Astronomy Observatory is operated by Associated
Universities, Inc., under cooperative agreement with the National Science
Foundation}

The data were reduced and calibrated with standard continuum and spectral-line
VLBI techniques in AIPS\footnote{Astronomical Image Processing System}. The
amplitude calibration for the two experiments included corrections for
atmospheric opacity. Antenna gain curves and measurements of system temperature
were used to calibrate amplitude data for each of the VLBA stations. Amplitude
calibration for the VLA was based on VLA scans of 3C\,286, for which we adopted a
$22$\,GHz flux density of $2.55$\,Jy, and on bootstrapping of flux densities for
VLBI calibrators. To minimize systematic errors due to atmospheric opacity
effects, only the VLA scans of VLBI calibrators obtained near the elevation of
3C\,286 were used in bootstrapping. The data in each polarization were corrected
for the effect of time variable source parallactic angle, and the station
positions were corrected for motions due to plate tectonics by utilizing the US
Naval Reference Frame solutions 9810 and 9513.

In the $5$, $8$, and $15$\,GHz data, the residual delays and fringe rates due to
the troposphere and clock uncertainties were removed via observations of
calibrators 4C\,39.25, 3C\,273, OJ\,287, and OQ\,208. The final images of the
nuclear region of NGC 3079 were obtained through application of self-calibration.
In the spectral data, the complex bandpass shape and any electronic phase
difference between the IF bands were removed by observation of calibrators
4C\,39.25, 1150+812, 1308+326, and 3C\,345. The position of the maser was
corrected by a fringe rate analysis of 10 spectral channels including the
brightest spectral feature at $956$\,km\,s$^{-1}$ with the result
$\alpha_{2000}=10^{h}01^{m}57\fs802\pm0.001$, $\delta_{2000}=55\degr
40\arcmin47\farcs26\pm0.01$. Phase and amplitude fluctuations due to troposphere
and clock uncertainties were removed by self-calibration of the brightest maser
feature at $956$\,km\,s$^{-1}$ and application of the resulting solution to all
spectral channels in all IF bands. After calibrating and imaging the
spectral-line data set, the relative positions of the maser features were
obtained by fitting two-dimensional elliptical Gaussians to the distribution of
the maser emission in each spectral channel. We obtained a continuum image at
$22$\,GHz by averaging all line-free frequency channels. Table \ref{tab2} lists
beam dimensions and noise levels at each frequency. Table \ref{tab1} lists the
properties of all spectral-line observations of the water maser with the VLBA.
The experiment presented in this study images the widest velocity range of
emission among all VLBA observations made to-date.

\begin{table}[!h]
\begin{center}
\caption{Half-power beam dimensions and noise levels.\label{tab2}}
\begin{tabular}{cccccc}\tableline\tableline
Frequency & Major & Minor & P.A.
& RMS \\ 
(GHz) & (mas) & (mas) & ($\degr$) & (mJy beam$^{-1}$)  \\ \tableline
5 & $3.6$ & $2.4$ & $14$ & $0.060$ \\
8 & $2.0$ & $1.3$ & $29$ & $0.055$ \\
15& $2.0$ & $1.3$ & $-89$ & $0.11$ \\
22& $0.30$ & $0.26$ & $-43$ &
$0.14$\tablenotemark{a},$\,\,\,\sim 2.3$\tablenotemark{b} \\
\tableline
\end{tabular}

\tablenotetext{a}{continuum} \tablenotetext{b}{spectral-line for a channel width
of $0.42$\,km\,s$^{-1}$.}

\end{center}
\end{table}

\begin{table}[!h]
\begin{center}
\caption{VLBA studies of the water maser in NGC 3079.\label{tab1}}
\begin{tabular}{lccccc}\tableline\tableline
& Trotter et al. 1998 & Sawada-Satoh et al. 2000 & This study \\
\tableline Date & 1995 Jan 9 & 1996 Oct 20 & 2001 Mar 23 \\
$V$ Range\tablenotemark{a} & $880-1204$ & $590-1008$ & $726-1491$ \\
$\Delta V$\tablenotemark{b} & $0.21$ & $0.42$  & $0.42$ \\
$\sigma$\tablenotemark{c} & $0.22$ & $0.54$ & $0.14$ \\
$\sigma$\tablenotemark{d} & $2.8$  & $6.6$ &  $2.3$  \\
\tableline
\end{tabular}

\tablenotetext{a}{Total velocity coverage in km\,s$^{-1}$}

\tablenotetext{b}{Channel spacing in km\,s$^{-1}$}

\tablenotetext{c}{The rms noise level in the pseudo continuum image in Janskys}

\tablenotetext{d}{The rms noise level in the spectral-line image in Janskys}

\end{center}
\end{table}

\begin{figure*}[!th]
        \centerline{\includegraphics[height=5in]{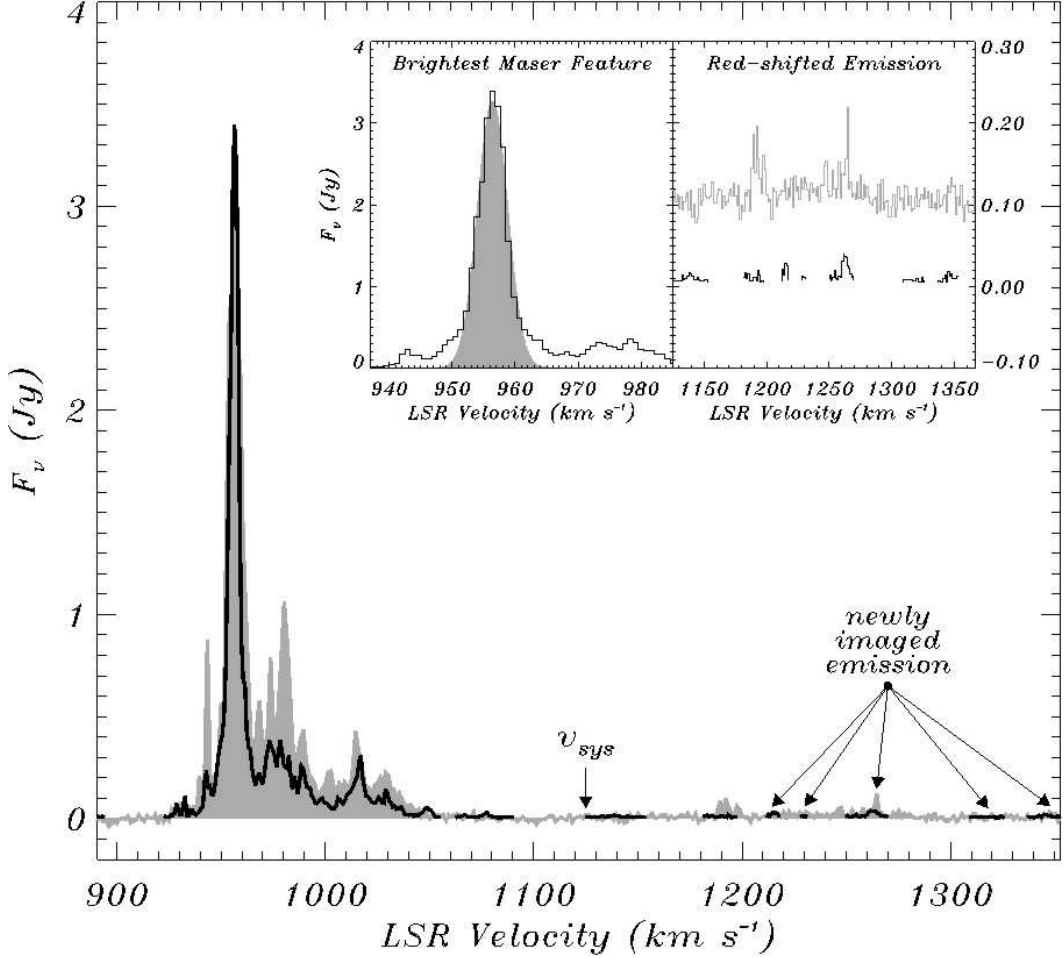}}
        \figcaption{Imaged-power and total-power spectra of the maser in NGC 3079.
        The total-power spectrum (gray)
        was obtained by position-switching with a 70-m Deep Space Network antenna in
        Robledo, Spain on 2003 February 4.
        The spectrum of the imaged power (black line) is based on the VLBI observation conducted
        on 2001 March 23 and was computed by summing the flux densities of the
        individual maser spots. {\it Left inset:} Imaged-power spectrum (black line) of
        the brightest maser feature at $\sim956$\,km\,s$^{-1}$ and a
        Gaussian fit to the main peak (gray) corresponding to an isotropic luminosity
        of $131$\,$L_{\sun}$. {\it Right inset:} Total-power and imaged-power spectra of
        the red-shifted emission with the vertical scale
        expanded and the total-power spectrum shifted upward to facilitate comparison.
        The vertical arrow indicates the systemic velocity of the galaxy ($1125$\,km\,s$^{-1}$).\label{spectrum}}
        \hrulefill\
\end{figure*}

To estimate the fraction of the power imaged by the interferometer, we acquired a
single-dish spectrum of NGC 3079 (see Fig.\ref{spectrum}) by position-switching
in a single polarization mode with the NASA Deep Space Network 70-m antenna in
Robledo, Spain, on 2003 February 4, using an observing setup similar to that
described in \cite{Greenhill2003survey}. After integrating for a total of
$2.5$\,hours, we achieved an rms noise level in the spectrum of $11$\,mJy for
$1.3$\,km\,s$^{-1}$ channel spacing. The antenna gain curve was obtained in a
single track by measurement of the elevation dependence of 1308+326 antenna
temperature, corrected for atmospheric opacity estimated from a tipping scan. The
antenna efficiency was estimated to be $0.43\pm0.01$ based on several
measurements of antenna temperature for 3C286. From the rms deviation of our gain
measurements about the best fit polynomial and the formal uncertainty in the
efficiency, we estimate that the gain calibration of the antenna is accurate to
within $10$\%.

\begin{figure*}[!th]
        \centerline{\includegraphics[height=5in]{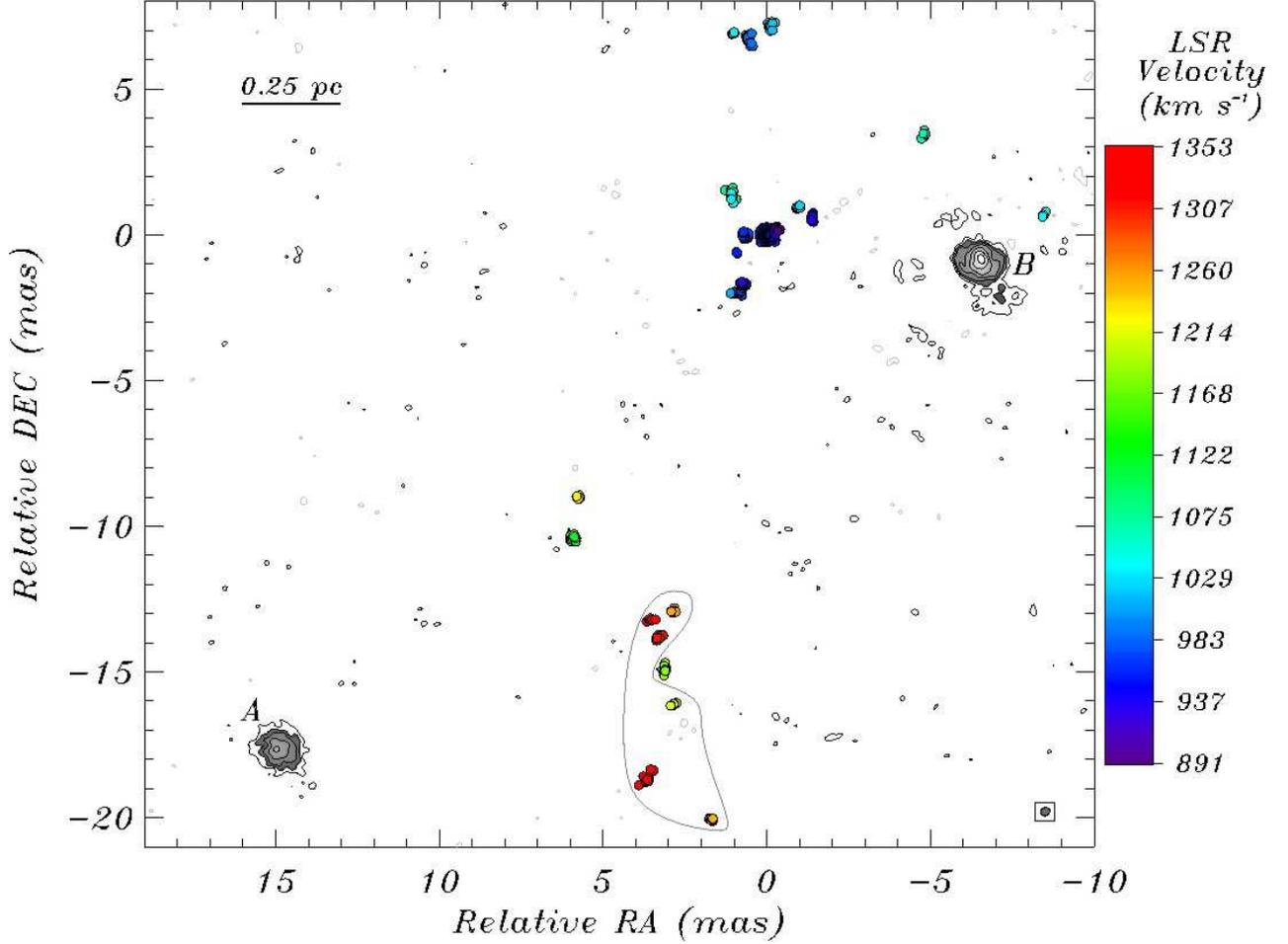}}
        \figcaption{Distribution of maser emission (colored circles)
        and of the $22$\,GHz continuum (gray contours) in the nuclear region of
        NGC 3079. The color of the maser spots indicates
        line-of-sight velocity in accordance with
        the bar on the right. The continuum contour levels are $-3\sigma,\,3\sigma,\,5\sigma \times 2^{k/2}$, where $\sigma=0.14$\,mJy\,beam$^{-1}$ and
        $k=0,1,2,3,\dots$. Contours above $5\sigma$ are shaded.
        The gray curve encloses the newly imaged emission, while the ellipse
        in the bottom right
        corner of the figure
        illustrates the resolution beam.\label{maser}}
        \hrulefill\
\end{figure*}

\begin{figure*}[!th]
        \centerline{\includegraphics[width=6.4in]{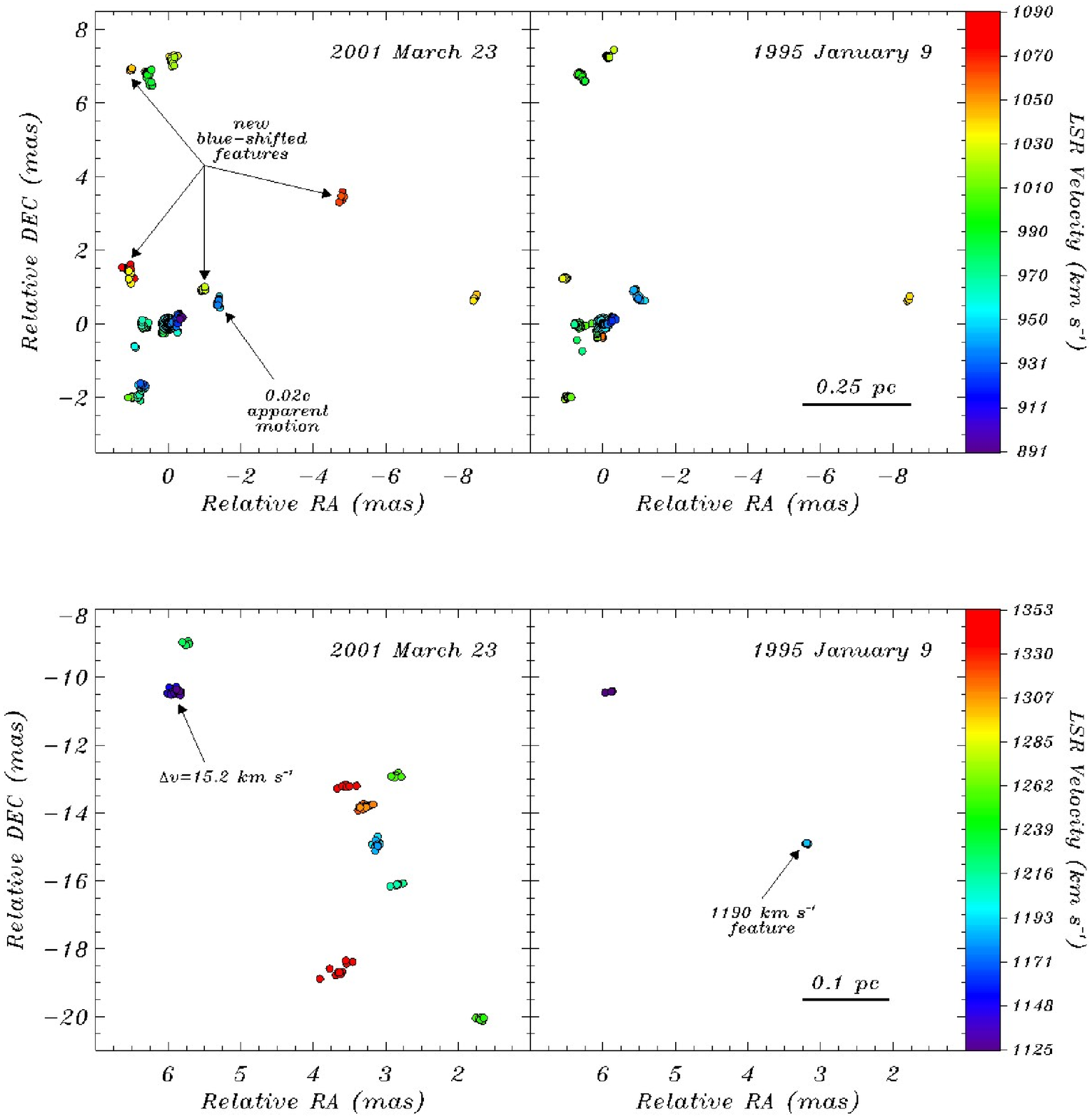}}
        \figcaption{Angular distribution of the water maser emission in NGC 3079 obtained on 1995 January 9
        \citep{Trotter1998} and on 2001 March 23 (this study). The top and bottom panels show emission blue- and
        red-shifted with respect to the systemic velocity, $1125$\,km\,s$^{-1}$, respectively.
        The velocity and angular scales are expanded
        to facilitate comparison between the two epochs.\label{maser_zoom}}
        \hrulefill\
\end{figure*}

\section{Results}
\label{results}
\subsection{Maser}
In agreement with previous observations of NGC 3079, we found that the maser
exhibits a conspicuous asymmetry whereby the blue-shifted emission significantly
dominates the detected flux density (Fig. \ref{spectrum}). The spectrum of the
imaged power agrees satisfactorily with the spectrum detected with the Effelsberg
100-m antenna 10\,days prior to the VLBI observation \citep[see Fig.2c in
][]{Hagiwara2002}. The observation with the NASA Deep Space Network 70-m antenna
in Robledo, Spain, conducted 1.9\,years later showed significant flux variability
throughout the spectrum except for the strongest feature (Fig. \ref{spectrum}).
Such variations, however, are not unusual since single-dish monitoring of the
maser has revealed substantial flux variability on timescales of years
\citep{Nakai1995, Baan1996, Hagiwara2002}.

As with the previous investigations \citep{Trotter1998, Sawada-Satoh2000}, we
found that the maser emission is distributed in a disordered linear structure at
P.A.$\sim -10\degr$ (Fig. \ref{maser}). Furthermore, the distribution of emission
on the sky is clearly segregated by velocity with emission blue- and red-shifted
with respect to the systemic velocity located in the northern and southern
section of the image, respectively. To determine temporal changes in the maser
distribution, we compared the maser positions obtained here with those of
\cite{Trotter1998}. Both the 1995 January 9 \citep{Trotter1998} and the 2001
March 23 (this study) epochs were calibrated by referencing the phases of all
spectral channels to the brightest maser feature at $\sim 956$\,km\,s$^{-1}$ and
thereby are most likely registered to the same volume of masing gas. Indeed,
three-dimensional flux density-weighted correlation of the distribution of the
maser emission presented in this study with that presented in \cite{Trotter1998}
indicates that the two epochs are aligned to within $\sigma=3.2$\,km\,s$^{-1}$ in
velocity and are registered by the self-calibration technique to within
$\sigma=0.5$\,mas on the sky ($0.5$\,mas corresponds to a transverse speed of
$0.02\,c$ over $6.25$\,years and is not an intrinsic measurement limit on the
maser proper motions). In fact, spectra of nearly half of the maser emission
clumps imaged in both \cite{Trotter1998} and this study peak within
3.2\,km\,s$^{-1}$ of each other, which is comparable to the
$\sigma=2.4$\,km\,s$^{-1}$ half width of the brightest maser feature (Fig.
\ref{spectrum}). The comparison of the two epochs (Fig. \ref{maser_zoom})
indicates that {(1)} the newly imaged red-shifted emission is confined to a
region south to south-west from the previously known blue-shifted features, {(2)}
several new blue-shifted features have appeared, {(3)} the maser feature located
at $(5.9,-10.5)$\,mas appears to have drifted in velocity from 1123 to
$1139$\,km\,s$^{-1}$, a change of $\sim16$\,km\,s$^{-1}$ in 6.25\,years, and
{(4)} the maser feature located at velocity $\sim940$\,km\,s$^{-1}$ and at
position $(-1.4,0.7)$\,mas appears to have moved westward with velocity of
$\sim0.02\,c$ ($\sim6000$\,km\,s$^{-1}$). The latter two cases might each be due
to variability of separate and distinct gas clumps rather than physical motions.
Note that the maser feature that has drifted in velocity between
\cite{Trotter1998} and this study is significantly displaced on the sky from the
$\sim1190$\,km\,s$^{-1}$ maser feature [located at ($3.1,-14.9$)\,mas], on whose
velocity drift the model of \cite{Sawada-Satoh2000} depends. Based on the
\cite{Trotter1998} and \cite{Hagiwara2002} studies, we place an upper limit of
$0.5$\,km\,s$^{-1}$\,yr$^{-1}$ on the velocity drift of the $\sim
1190$\,km\,s$^{-1}$ feature.

\begin{figure*}[!th]
        \centerline{\includegraphics[height=5in]{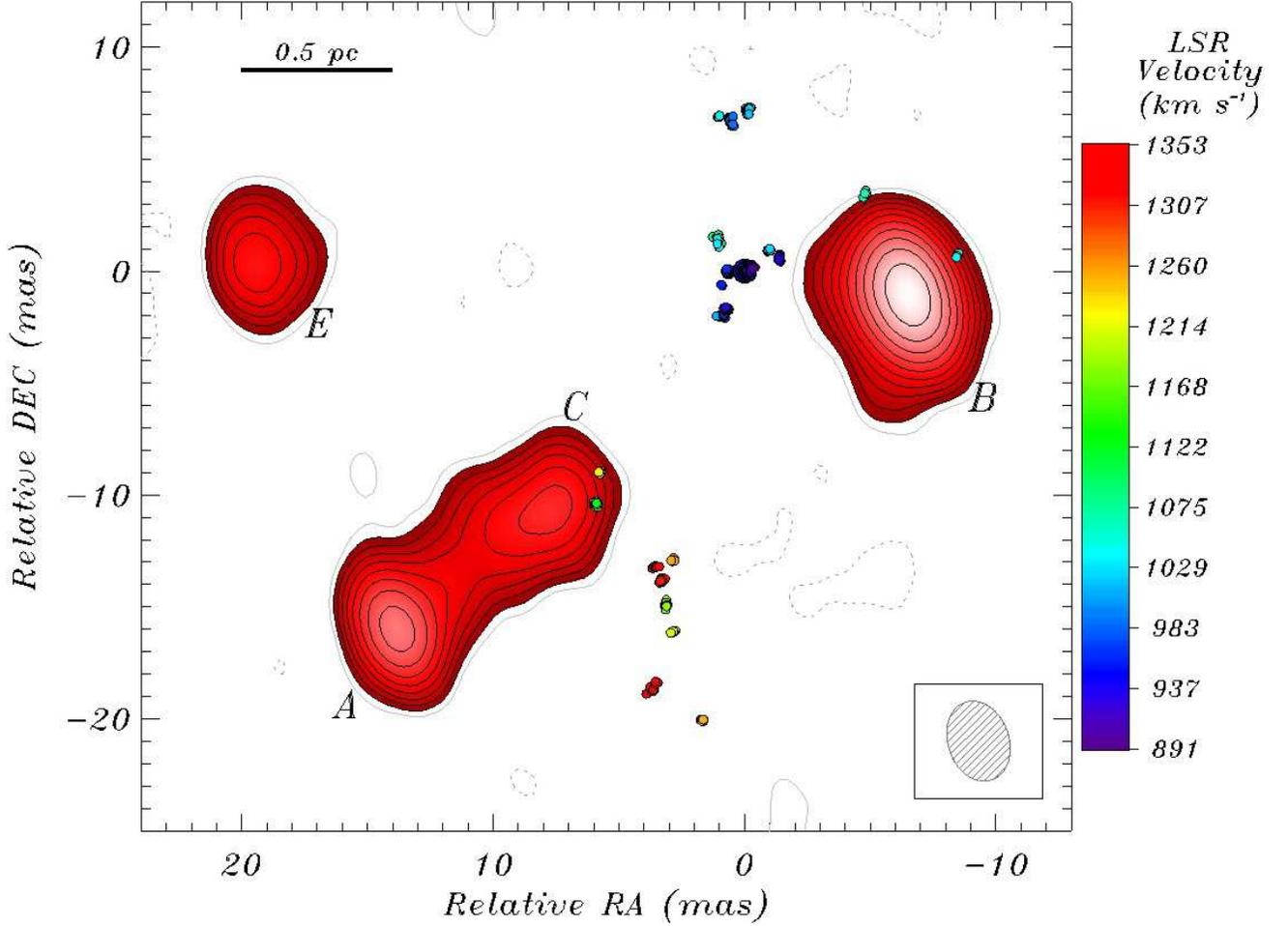}}
        \figcaption{Contours of the 5\,GHz continuum drawn at levels
        $-2\sigma,\, 4\sigma,\,6\sigma \times 2^{k/2}$, where $\sigma=0.060$\,mJy\,beam$^{-1}$ and
        $k=0,1,2,3,\dots$. Contours above $6\sigma$ are shaded. Compact component D detected by \cite{Trotter1998}
        at 5\,GHz with flux density of $4.3\pm 0.5$\,mJy would be located at $(38.3,-31.9)$\,mas in the above
        plot but was not detected at
        a level of $6\sigma=0.36$\,mJy.
        The maser features shown here as small filled circles were registered
        to the 5\,GHz continuum by aligning the positions of component B at 5 and 22\,GHz.
        The uncertainties in this registration are $0.02$ and $0.03$\,mas in right ascension
        and declination, respectively. The conversion from the color of the maser spots
        to their line-of-sight velocity is given by the bar on the right.
        Beam dimensions are shown in the bottom right corner.\label{5GHz}}
        \hrulefill\
\end{figure*}

\begin{figure*}[!th]
        \centerline{\includegraphics[height=5in]{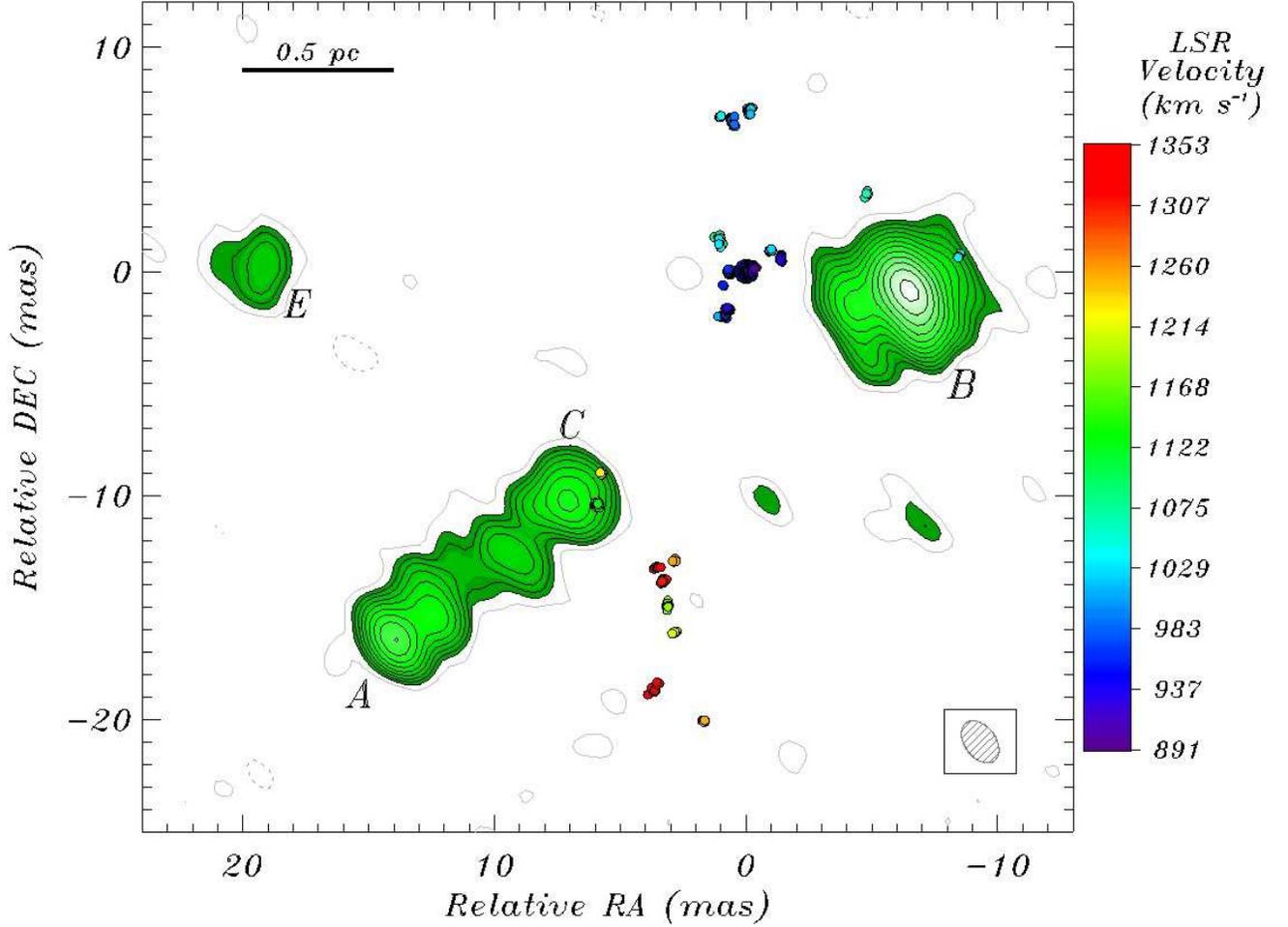}}
        \figcaption{Contours of the 8\,GHz continuum drawn at levels
        $-3\sigma,\,3\sigma,\,6\sigma \times 2^{k/2}$, where $\sigma=0.055$\,mJy\,beam$^{-1}$ and
        $k=0,1,2,3,\dots$. Contours above $6\sigma$ are shaded. The maser
        features are shown using small filled circles color-coded
        by velocity in accordance with the bar on the right. The
        registration of the maser to the 8\,GHz continuum was achieved
        by aligning the positions of component B at 8 and 22\,GHz.
        The uncertainty in this registration is $0.01$\,mas in both right ascension
        and declination.
        The ellipse in the bottom right corner of the figure illustrates beam dimensions.\label{8GHz}}
        \hrulefill\
\end{figure*}

\begin{figure*}[!th]
        \centerline{\includegraphics[height=5in]{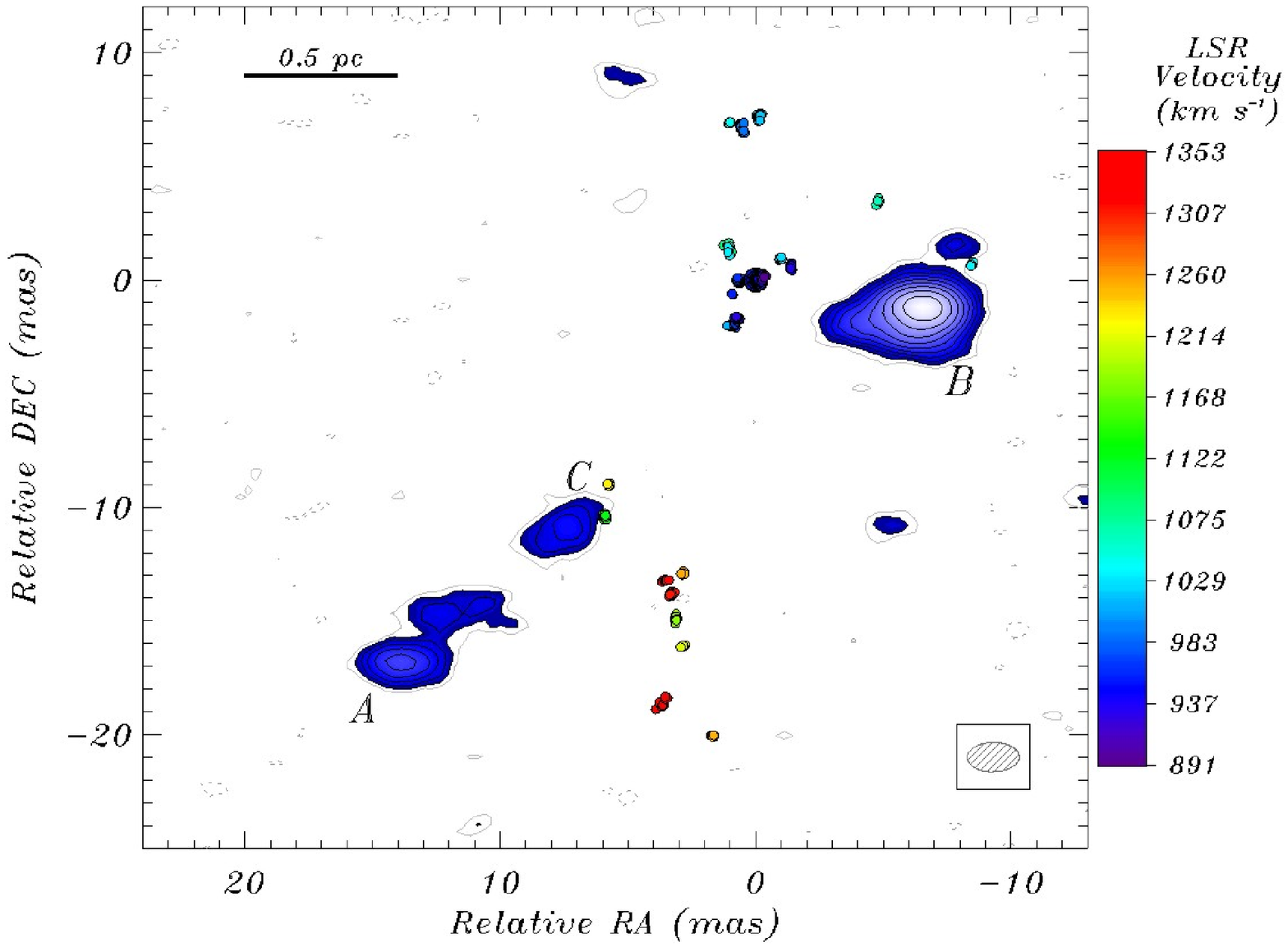}}
        \figcaption{Contours of the 15\,GHz continuum drawn at levels
        $-2\sigma,\,4\sigma,\,6\sigma \times 2^{k/2}$, where $\sigma=0.11$\,mJy\,beam$^{-1}$ and
        $k=0,1,2,3,\dots$. Contours above $6\sigma$ are shaded. Superposed
        on top of the continuum are the maser features registered to the
        15\,GHz continuum by aligning the positions of component B at 15 and 22\,GHz.
        The uncertainty in this registration is $0.01$\,mas in both right
        ascension and declination
        directions. The bar on the
        right indicates the relationship between the color and the velocity of the maser features.
        Beam dimensions are shown in the bottom right corner of the figure.\label{15GHz}}
        \hrulefill\
\end{figure*}

\subsection{Continuum}
\label{continuum} 
\begin{figure*}[!th]
        \centerline{\includegraphics[width=5.5in]{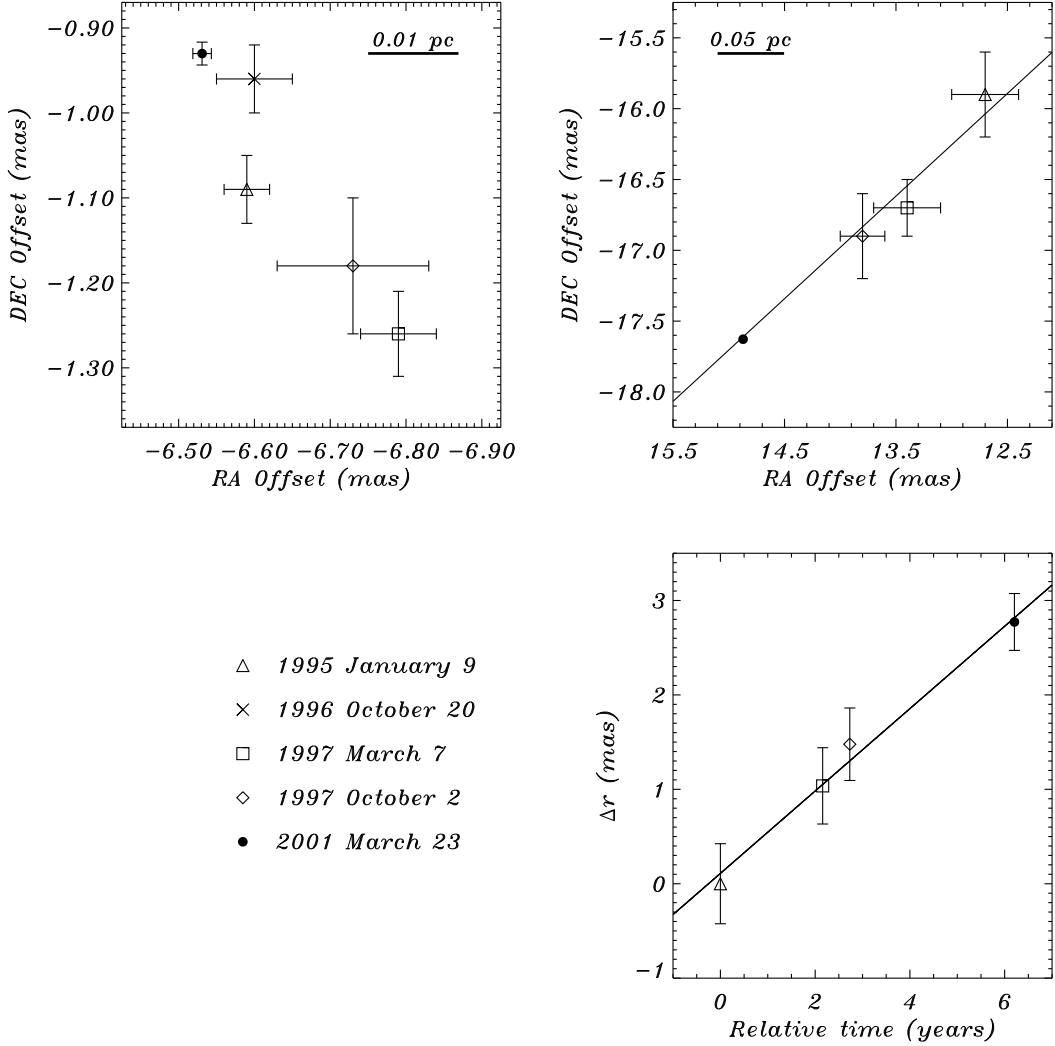}}
        \figcaption{{\it Left:} Position of
        component B with respect to the reference maser feature.
        The small scatter (typical magnitude of $\sim 0.35$\,mas) in the
        relative position of component B
        is not consistent with formal measurement uncertainties and
        might be due to morphological or positional changes
        in the reference maser feature or in component B itself.
        {\it Top right:} Position of
        A with respect to the reference maser feature.
        A fit to the four measured positions indicates that A is moving
        along P.A. $\sim 126\degr\pm 5\degr$, effectively away from component B.
        {\it Bottom right:} Temporal motion of
        A along the linear trajectory in the top right panel, consistent
        with a transverse
        speed of $0.12\pm 0.02\,c$.
        The 1995 January 9 epoch was obtained
        from \cite{Trotter1998}, epochs
        1996 October 20, 1997 March 7, and 1997 October 2 are from \cite{Sawada-Satoh2002},
        while the 2001 March 23 epoch is from this study.
        For points where error bars are not visible, the errors are smaller than the
        plotting symbol.\label{motion}}
        \hrulefill\
\end{figure*}

Components A and B were detected at all frequencies while
component C was detected at 5, 8, and 15\,GHz, but not at 22\,GHz (Fig.
\ref{5GHz}, \ref{8GHz}, \ref{15GHz}, and \ref{maser}). In agreement with previous
reports, components A, B, and C are collinear along the position angle of $\sim
126\degr$. To estimate the flux density, position, and position uncertainty of
each continuum component, we fitted multiple two-dimensional Gaussian model
components to the emission peaks at each frequency (Table \ref{tab3}). In
agreement with previous reports \citep[e.g.,][]{Sawada-Satoh2000} and
supplementing our data with published positions, we found that the systematic
motion of component A at 22\,GHz is consistent with a transverse speed of
$0.12\pm 0.02\,c$ along P.A. of $\sim 126\degr\pm 5\degr$, which is approximately
away from component B (Fig.\ref{motion}). Note that the change (typical magnitude
of $\sim0.35$\,mas) from epoch to epoch in the position of component B with
respect to the reference maser feature does not appear to be systematic and is
inconsistent with formal measurement uncertainties (Fig. \ref{motion}). This
observed scatter might be due to the change in morphology or position of either
component B or the reference maser feature. However, the time average position of
component B is approximately stationary with respect to the maser emission. We
emphasize that the jitter of component B, like the motion of E in the direction
of B reported by \cite{Middelberg2003}, might not be a real but an apparent
motion due to changing source characteristics in a flow or across shocks (e.g.,
Kellermann et al. 2004).

\begin{table}[!h]
\begin{center}
\caption{Characteristics of continuum components at 5, 8, 15, and 22\,GHz
obtained by two-dimensional multiple component Gaussian fits to emission
features. \label{tab3}}\footnotesize
\begin{tabular}{cccccccc}\tableline\tableline
Frequency & Component & $S_{\nu}$ & $R$ & $\phi$ &
Major & Minor  & P.A.\\
(GHz) & & (mJy) & (mas) & ($\degr$) & (mas) & (mas) & ($\degr$) \\
\tableline
5 & B & $13.4 \pm 0.4$ &  --- & --- & $3.63\pm 0.07$ & $1.62\pm 0.03$ & $19\pm 1$  \\
  & A & $4.7\pm 0.3$ & $25.25 \pm 0.05$ & $126.5 \pm 0.1$  & $3.0\pm 0.1$ & $1.60\pm 0.07$ & $20\pm 3$ \\
  & C & $3.6 \pm 0.5$& $17.8 \pm 0.1$ & $123.5 \pm 0.4$ & $3.4\pm 0.4$ & $2.7\pm 0.3$ & $109\pm 19$ \\
  & E & $2.8\pm 0.6$ &  $25.9 \pm 0.2$ & $85.5 \pm 0.5$ & $4.0 \pm 0.7$ & $2.7\pm 0.4$ & $32\pm 17$
  \\ \tableline
8 & B & $32.4 \pm 0.2$ & --- & --- & $2.89\pm 0.01$ & $1.87 \pm 0.01$ & $24.2\pm 0.3$  \\
  & A & $11.5\pm 0.2$ & $25.07 \pm 0.02$ & $126.94 \pm 0.03$ & $3.11\pm 0.04$ & $2.31 \pm 0.03$ & $140\pm 2$  \\
  & C & $9.7 \pm 0.3$ & $16.81 \pm 0.03$ & $124.4 \pm 0.1$ & $4.5\pm 0.1$ & $2.58\pm 0.06$ & $135\pm 2$  \\
  & E & $2.5 \pm 0.2$ &  $25.90 \pm 0.06$ & $87.4 \pm 0.2$ & $2.9 \pm 0.2$ & $2.5 \pm 0.2$ & $175\pm 20$
  \\ \tableline
15& B & $27.3 \pm 0.3$ & --- & --- & $2.71 \pm 0.02$ & $1.95\pm 0.02$ & $97.6 \pm 0.6$\\
  & A & $7.8\pm 0.5$ &  $24.84 \pm 0.09$ & $127.2 \pm 0.2$ & $4.6\pm 0.3$ & $2.3\pm 0.1$ & $128\pm 3$ \\
  & C & $3.7 \pm 0.4$ & $16.91 \pm 0.09$ & $124.6 \pm 0.3$ & $3.0\pm 0.2$ & $2.0\pm 0.2$ & $130\pm 7$
  \\ \tableline
22& B & $40 \pm 1$ & --- & --- & $1.09 \pm 0.03$ & $0.88\pm 0.03$ & $31 \pm 6$\\
  & A & $28 \pm 2$ &  $27.14 \pm 0.03$ & $127.96 \pm 0.06$ & $1.24\pm 0.06$ & $1.08\pm 0.05$ & $24\pm 14$ \\\tableline
\end{tabular}

\tablecomments{Angular separation $R$ and position angle $\phi$ are measured with
respect to component B at each frequency. The 5, 8, and 15\,GHz data were
acquired on 1996 February 12 while the 22\,GHz data were obtained on 2001 March
23. Component D ($S_{\nu}=4.3\pm0.5$\,mJy) at position $r=54.5\pm0.4$\,mas and
$\phi=124.7\degr\pm0.4\degr$ and detected by \cite{Trotter1998} was not detected
at a level of $6 \sigma=0.36$\,mJy in this study. One mas corresponds to
$0.084$\,pc.}

\end{center}
\end{table}

\begin{figure*}[!th]
        \centerline{\includegraphics[height=5in]{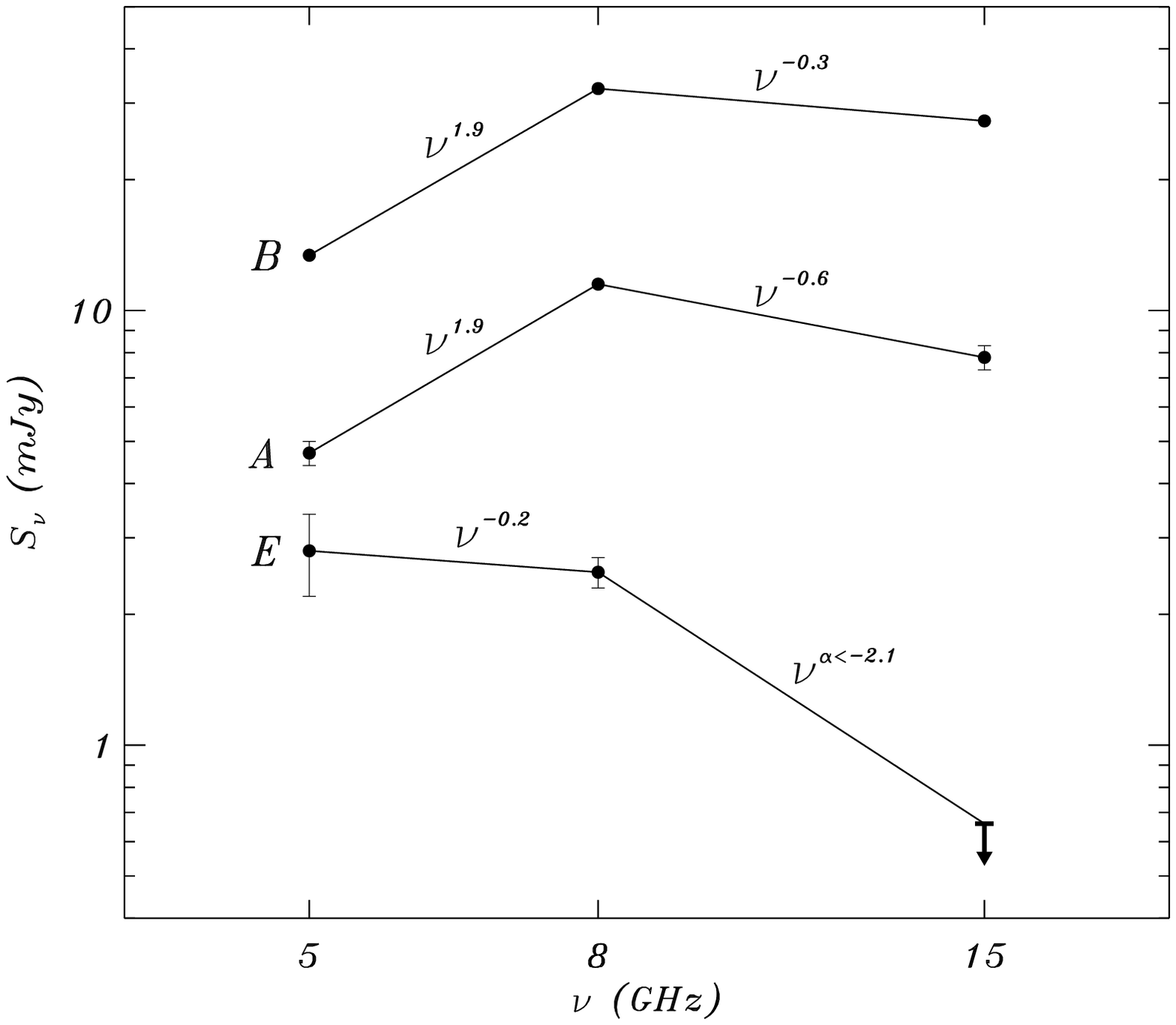}}
        \figcaption{Spectra of continuum components A, B, and E and corresponding spectral
        indices $\alpha$, where $S_{\nu}\propto \nu^{\alpha}$.
        The arrow indicates an upper limit on the flux density of component E at 15\,GHz.
        The effect of resolution on the spectral index
        is probably small because the beam dimensions at the three frequencies are
        comparable (Table \ref{tab2}).
        Flux densities of the continuum components at $22$\,GHz are not shown
        because they were obtained at a different epoch, and the emission is known to
        be time variable.\label{spectral_index}}
        \hrulefill\
\end{figure*}

We confirmed the spectral turnover reported by \cite{Trotter1998} of components A
and B and constrain the frequency of this turnover to the range between 5 and
15\,GHz (Fig. \ref{spectral_index}). To place more stringent constraints on the
turnover frequencies, we considered two simple spectral models for each
component. In the first model, we assumed that at lower frequencies the spectra
are dominated by synchrotron self-absorption ($S_{\nu}\propto \nu^{5/2}$)
intrinsic to the source, while at higher frequencies the spectra are determined
by optically-thin synchrotron emission ($S_{\nu}\propto \nu^{\alpha}$) also
internal to the source. Under these assumptions, the spectra of components A and
B can be characterized by $S_{\nu}=a\nu^{5/2}(1-e^{-b\nu^{\alpha-5/2}})$, where
$a$, $b$, and $\alpha$ are parameters of the model. The flux densities measured
at three frequencies for each component uniquely determine the three parameters
of the model, which yielded $\alpha=-1.5$ and $-1.1$, and turnover frequencies
(i.e., frequency of maximum flux density) of $9.0$ and $9.6$\,GHz for components
A and B, respectively. For the data reported by \cite{Trotter1998}, we obtained
$\alpha=-1.5$ and $-1.3$, and turnover frequencies of $9.1$ and $9.0$\,GHz for
components A and B, respectively. However, if free-free absorption
($\tau_{\nu}\propto \nu^{-2.1}$) extrinsic to the source dominates at lower
frequencies instead of synchrotron self-absorption, then the spectra of
components A and B can be modelled by $S_{\nu}=a\nu^{\alpha}e^{-\beta\nu^{-2.1}}$
\citep*[e.g.,][]{Bicknell1997,Tingay2001}. For this model, we obtained
$\alpha=-1.8$ and $-1.3$ as well as turnover frequencies of $8.7$ and $9.5$\,GHz
for components A and B, respectively. Similarly, the flux densities reported by
\cite{Trotter1998} yielded $\alpha=-1.8$ and $-1.6$ as well as peak locations of
$8.8$ and $8.7$\,GHz. Thus, for both models and for both data sets, spectral
turnover frequencies lie in the range $8-10$\,GHz.

The spectra of components A and B are reminiscent of spectra of gigahertz peaked
spectrum (GPS) sources, as already noted by \cite{Trotter1998}. These sources are
thought to arise from regions where jets interact with the dense ambient gas
\citep[e.g.,][]{Bicknell1997}. Their spectra peak in the frequency range of
$0.1-10$\,GHz with high-frequency spectral indices of
$-1.3\lesssim\alpha\lesssim-0.5$ and low-frequency indices of $\alpha\sim1$.
Because of the similarity between the two components and the GPS sources, we
suggest that components A and B arise in regions where a relativistic outflow is
interacting with a dense ambient medium. Since components A and B consist of
multiple components and do not exhibit a flat spectrum, neither A nor B is likely
to be a jet core \citep[cf.][]{Sawada-Satoh2000}.

\cite{Trotter1998} reported a detection at $5$\,GHz with a signal-to-noise ratio
(S/N) of $39$ of a compact $4.3\pm 0.5$\,mJy component (D) located $\sim 4$ pc
southeast of B and collinear with components A, B, and C. However, we found no
evidence above the $0.36$\,mJy\,$(6\sigma)$ level of this component at 5\,GHz.
Since the 5\,GHz beam area of \cite{Trotter1998} was $2.4\times$ larger than the
5 GHz beam area in our study, the non-detection of D could be due to either
resolution effects or source variability. Component D was also not detected at
5\,GHz by \cite{Middelberg2003}.

The continuum images at 5 and 8\,GHz revealed a new component (E) that is not
collinear with previously known components A, B, C, and D (Fig. \ref{5GHz} and
Fig. \ref{8GHz}). We detected component E with an S/N of 31 and 16 at 5 and
8\,GHz, respectively \citep{Kondratko2000}. \cite{Middelberg2003} confirmed the
detection of E and also reported identification at 1.7 and 2.3\,GHz of a new
component F located roughly $3.7$\,pc from component B along P.A. of
$\sim100\degr$. Since the beam size in the \cite{Trotter1998} study was larger
than the beam size in our study, component E should have been detected by
\cite{Trotter1998} with an S/N of at least 17 at 5\,GHz if its flux density were
constant. The failure of \cite{Trotter1998} to detect E indicates that the source
is most likely time variable. Indeed, comparison of our results with those of
\cite{Trotter1998} reveals variation in the flux density of continuum component A
at $22$\,GHz by a factor of $\sim5$ over $6.2$\,years. If we compare the results
of \cite{Sawada-Satoh2000} and of this study, the variation in the flux density
of component A at $22$\,GHz could have been as large as a factor of $\sim52$ over
$4.4$\,years (we note that, although the $22$\,GHz beam of
\cite{Sawada-Satoh2000} is larger, they fail to detect component A). In both
cases, the timescale of variability is greater than the light travel time within
the components ($<2$ yr).

Our flux density estimates for component E indicate a spectral index of
$\alpha=-0.2$ between 5 and 8\,GHz and a spectral index of $\alpha<-2.1$ between
8 and 15\,GHz (Fig.\ref{spectral_index}). The high-frequency decline in the flux
density of component E is too steep to be consistent with optically-thin
synchrotron emission [$\alpha=(1-p)/2$], since it would require $p>5.2$, much
greater than the normally assumed value of $p\sim2.4$, where $p$ is the power law
index on the electron energy distribution [$N(E)\propto E^{-p}$]. However, for a
situation in which there is no continuous injection of electrons into the
synchrotron source and the pitch angle distribution remains conserved, the
cooling of high-energy electrons leads to a spectral index of $\alpha =
-(2p+1)/3$ above a break frequency $\nu_b=[B/(1 \mbox{ G})]^{-3} [t/(1\mbox{
yr})]^{-2}$\,GHz \citep{Kardashev1962}. Hence, nominally, a field of $\sim40$\,mG
would give $\nu_b\sim10$\,GHz after a cooling time of $\sim 40$\,years, the
travel time of component E at $0.12\,c$ from the central engine (refer to Section
\ref{center}). A value of $p>2.6$ would be required to give a spectral index of
$\alpha<-2.1$ above the break frequency, consistent with our observations. Using
the same magnetic field and extending the cooling time by $\sim 5.8$\,years
(i.e., the time difference between the two studies), we estimate a break
frequency at the observing epoch of \cite{Middelberg2003} to be $\sim8$\,GHz,
consistent with their detection of E at $5$\,GHz and non-detection of E at
$15$\,GHz. We note that their spectrum appears to be shifted to lower frequencies
because it peaks between $1.7$ and $5$\,GHz rather than between $5$ and $8$\,GHz
as in our study. This shift is consistent with an adiabatic expansion of the
rapidly cooling synchrotron source, for a source lifetime of $\sim40$\,years,
$p\sim2.6$, and the ratio of the turnover frequencies of $\sim 1.4$
\citep[corresponding to a shift of roughly $1.5$\,GHz; note that turnover
frequency is different from $\nu_b$; Eq. 13.27 of][]{Kellermann1988}.

\begin{figure*}[!th]
        \centerline{\includegraphics[height=5in]{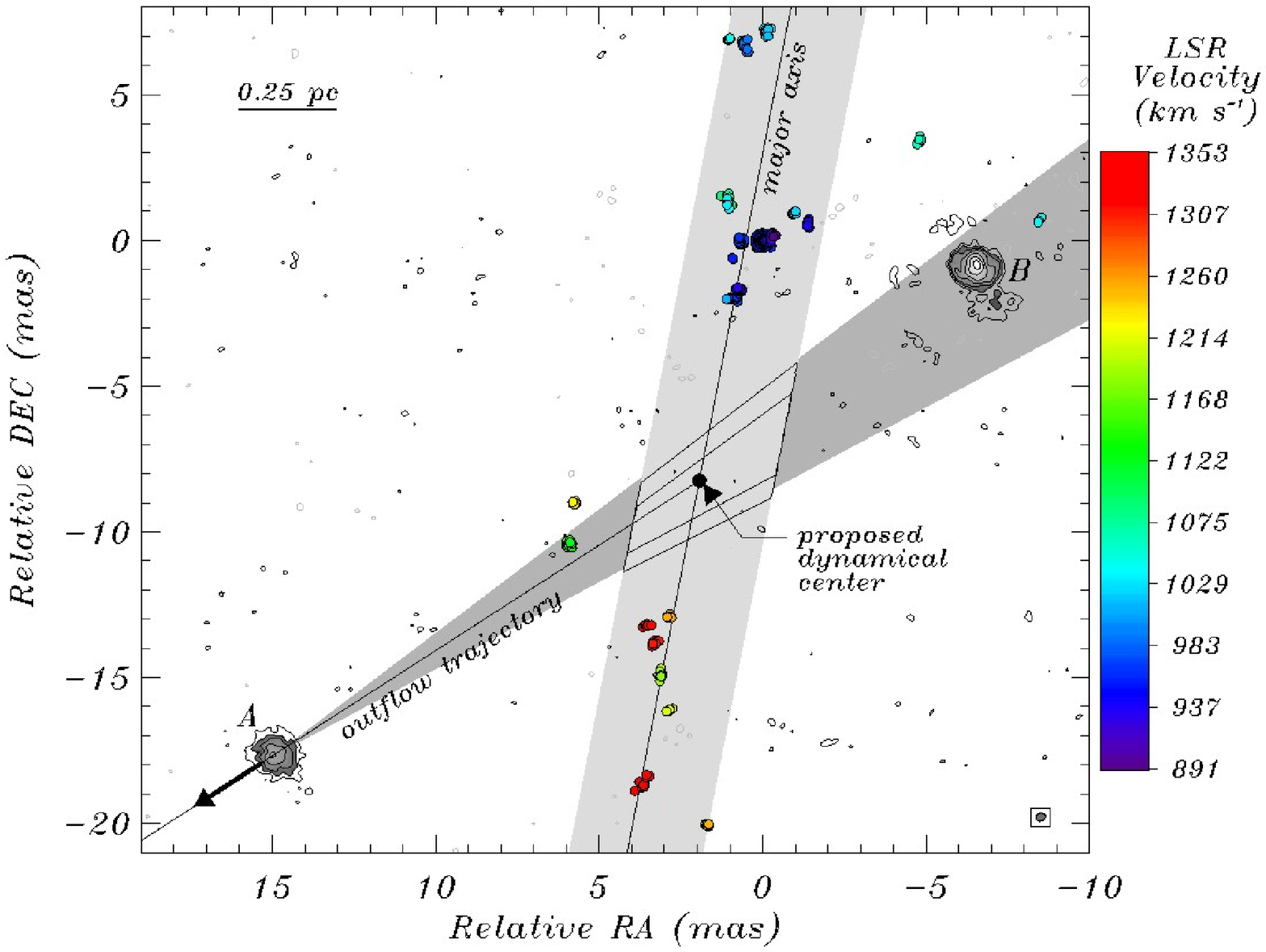}}
        \figcaption{Distribution of maser emission (colored circles)
        and of the $22$\,GHz continuum (gray contours) in the nuclear region of
        NGC 3079.
        The color of the maser spots indicates
        line-of-sight velocity in accordance with
        the bar on the right. The continuum contour levels
        are $-3\sigma,\,3\sigma,\,5\sigma \times 2^{k/2}$,
        where $\sigma=0.14$\,mJy\,beam$^{-1}$ and
        $k=0,1,2,3,\dots$. The nearly vertical line is the proposed
        disk major axis, while the gray nearly vertical region illustrates
        the uncertainty in the location and orientation
        of the axis.
        Also shown in the figure is the
        outflow trajectory of component A at
        P.A. of $126\degr\pm5\degr$ (as in Fig. \ref{motion}),
        where the cone shows the $1\sigma$ uncertainty in the
        orientation of this trajectory. The intersection of the
        outflow direction with the disk major axis yields the
        location for the dynamical center (filled symbol), where
        $70\%$ and $45\%$ confidence regions for
        its location are illustrated by the central contours.
        Note that the two easternmost and two westernmost maser
        features (outside the gray area depicting the maser disk)
        are most likely associated with the outflow.
        The ellipse in the bottom right
        corner of the figure
        illustrates beam dimensions.\label{maser2}}
        \hrulefill\
\end{figure*}

\begin{figure*}[!th]
        \centerline{\includegraphics[width=3.5in]{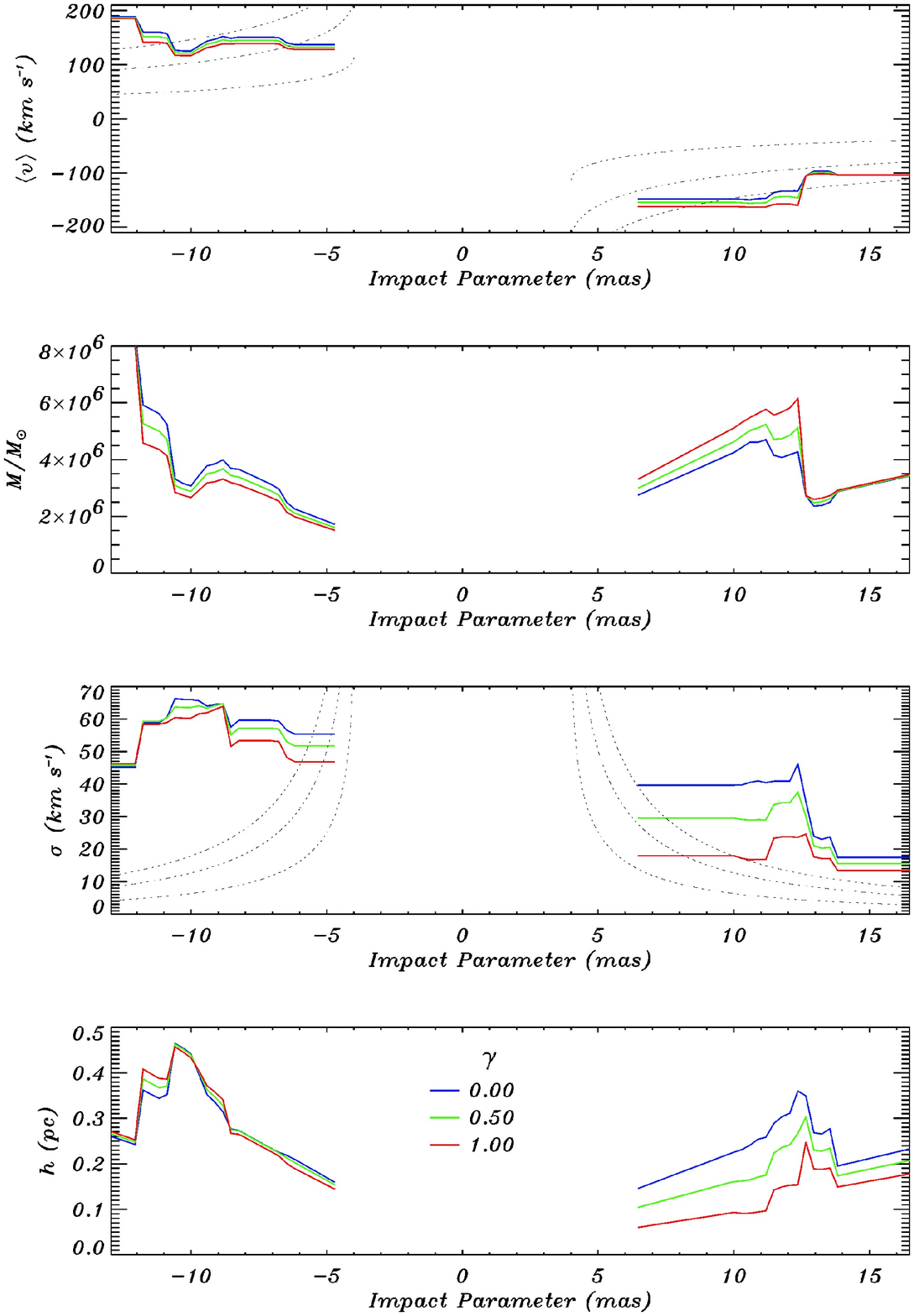}}
        \figcaption{Average velocity $\langle v\rangle$, enclosed total mass $M(r)$,
        velocity dispersion $\sigma$, and
        scale height $h$ as a function of
        distance from the location of the dynamical center along the disk major
        axis, i.e., impact parameter. At each impact parameter, the four quantities under
        consideration were computed using an averaging aperture of radius $4$\,mas ($0.34$\,pc)
        and weighting $F^{\gamma}$ for various values
        of $\gamma$, where $F$ is the maser flux density and $\gamma$ is denoted by color.
        The average velocity, velocity dispersion, enclosed mass, and disk scale height
        were computed
using $\langle v\rangle=\left[\sum_i F_i^{\gamma}\right]^{-1}\,\sum_i
F_i^{\gamma}\,(v_i-v_{sys})$, $M(r)=\langle v\rangle^2r/G$,
$\sigma^2=\left[\sum_i F_i^{\gamma}\right]^{-1}\,\sum_i
F_i^{\gamma}\,(v_i-v_{sys}-\langle v\rangle)^2$, and $h\sim r\sigma/\langle
v\rangle$, respectively, where the sums are over all maser features located
within the averaging aperture and $r$ is the impact parameter. The curves were
computed in a
        quasi-continuous fashion to
        emphasize the uncertainty in the computed parameters due to
        discreteness in velocity and spatial sampling.
        The dotted lines show the analytical results for a Keplerian edge-on disk of
        negligible mass around
        a $[0.5, 2, 4]\times 10^6$\,$M_{\sun}$ point mass computed
        using an averaging aperture of $4$\,mas and under the
        assumption that
        the maser emission traces completely
        the midline of the putative disk.\label{sigma}}
        \hrulefill\
\end{figure*}

\section{Discussion}
\label{discussion} In what follows, we propose that the inner parsec in NGC 3079
contains a nearly edge-on, massive, thick, and flared disk. The disk orbits a
mass of $\sim2\times 10^6$\,$M_{\odot}$ enclosed within $0.4$\,pc, most likely a
supermassive black hole, and it is aligned with the kpc-scale molecular disk. The
disk is most likely self-gravitating, clumpy, and forming stars. Accretion onto
the black hole drives a jet that is misaligned with the disk rotation axis and is
interacting with the dense ambient medium. The presence of off-axis aging
synchrotron components may suggest that the jet changes orientation. The jet may
coexist with a pc-scale wide-angle outflow, which is inferred from the
observation of masers at high latitudes above the disk and which might be related
to the known kpc-scale superbubble.

\subsection{Analysis of Maser Kinematics}
\subsubsection{The Disk Major Axis and the Dynamical Center}
\label{center} The largely north-south linear distribution of the maser emission,
aligned with the kpc-scale molecular disk, as well as the segregation of the
blue- and the red-shifted emission on the sky are suggestive of a nearly edge-on
pc-scale molecular disk. A fit to the angular distribution of maser emission
yields the following disk major axis: $\alpha({\rm mas})=(-0.17\pm 0.06) \times
\delta({\rm mas}) + (0.5\pm 0.7)$\,mas, where $\alpha$ and $\delta$ are right
ascension and declination, respectively, measured with respect to the reference
maser feature (Fig. \ref{maser2}). The position angle of the proposed major axis
is $\sim-10\degr$, in close agreement with the position angle of the kpc-scale
molecular disk \citep[P.A.$\sim -11\degr$;][]{Koda2002}.

We use the estimated disk major axis and the systematic motion of component A to
establish the location for the dynamical center. The uncertainty in the location
and orientation of the disk major axis restricts the position of the dynamical
center to lie within the gray nearly vertical band in Fig.\ref{maser2}. The
systematic motion of component A (Fig.\ref{motion}), on the other hand, limits
the location of the dynamical center to P.A.$\sim-54\degr\pm 5\degr$ as measured
with respect to A and illustrated as a cone in Fig.\ref{maser2}. The intersection
of these two loci yields $(\alpha,\delta)=(1.9\pm 1,-8.2\pm 2)$ for the mean
position of the dynamical center (shown in Fig. \ref{maser2}). Note that the
proposed position of the dynamical center lies in close proximity to the
intersection of the disk major axis with the line joining components A and B,
which was expected since the trajectory of A is nearly parallel to the line
joining A and B. The position of the dynamical center proposed in this study is
similar to that proposed in \cite{Trotter1998} but significantly different from
that proposed by \cite{Sawada-Satoh2000} (Fig. \ref{comparison}).

\subsubsection{Disk Structure}
To study the disk structure traced by the maser emission, we computed the average
velocity $\langle v\rangle$, the enclosed total mass $M(r)$, the velocity
dispersion $\sigma$, and the disk scale height $h$ as a function of distance $r$
from the putative dynamical center along the major axis of the disk, i.e., the
impact parameter (Fig. \ref{sigma}). In this computation and the following
analysis, we explicitly assume that the maser emission traces the kinematics of
the underlying molecular material, as is indeed the case in the archetypal maser
galaxy NGC 4258. To compute the four average quantities at each impact parameter,
we used an averaging aperture of radius $4$\,mas ($0.34$\,pc) and weighting
function $F^{\gamma}$, where $F$ is the maser flux density and $0\le \gamma \le
1$. This particular weighting scheme was adopted to demonstrate that the general
conclusions inferred from this analysis are relatively insensitive to the
observed broad range of the maser flux density (Fig. \ref{spectrum}). Moreover,
at each impact parameter, the average velocity and velocity dispersion were
computed in accordance with $\langle v\rangle=\left[\sum_i
F_i^{\gamma}\right]^{-1}\,\sum_i F_i^{\gamma}\,(v_i-v_{sys})$ and
$\sigma^2=\left[\sum_i F_i^{\gamma}\right]^{-1}\,\sum_i
F_i^{\gamma}\,(v_i-v_{sys}-\langle v\rangle)^2$, respectively, where the sums are
over all maser features located within the averaging aperture. Assuming Keplerian
rotation and spherical symmetry, we estimated the enclosed mass as $M(r)=\langle
v\rangle^2r/G$, where $r$ is the impact parameter. Under the assumption of
hydrostatic equilibrium and axisymmetric potential, the disk scale height is
given by $h\sim r\sigma/\langle v\rangle$. Note that the two easternmost and two
westernmost maser features have been omitted from this analysis because they do
not appear to be associated with the disk (Fig. \ref{maser2}). Their inclusion
would not change our conclusions significantly.

\subsubsection{Rotation, the Enclosed Mass, and the Disk Mass}
\label{enclosed} The computed average velocity displays a red-blue asymmetry
about the adopted $v_{sys}$ and estimated dynamical center and is thus consistent
with rotation (Fig. \ref{sigma}). The Keplerian rotation law yields a mass of
$\sim2\times10^6\,M_{\sun}$ enclosed within $0.4$\,pc (5\,mas).
\cite{Trotter1998} reported a similar enclosed mass, $\sim10^6$\,$M_{\sun}$, from
consideration of the blue-shifted emission only. In our case, however, both the
blue- and the red-shifted emission provide independent and consistent estimates
of enclosed mass. We note that since the maser features may lie somewhat away
from the disk midline or the disk might be somewhat tilted from edge-on, the
computed enclosed mass is a lower limit, but is probably correct to within
factors of order unity. For a central mass of $2\times10^6\,M_{\sun}$ and an
inner radius of $0.4$\,pc, the maximum centripetal acceleration (i.e., velocity
drift) and the maximum proper motion would be $0.05$\,km\,s$^{-1}$\,yr$^{-1}$ and
$2$\,$\mu$as\,yr$^{-1}$, respectively. The much larger apparent velocity drifts
and motions (Fig. \ref{maser_zoom}) are most likely due to variability of
separate gas clumps and macroscopic random motions among clumps rather than
global kinematics, which makes a measurement of velocity drifts and proper
motions due to the gravitational potential of a central mass difficult if not
impossible.

Since the rotation curve of both blue- and red-shifted features is flat, i.e.,
$M(r)\propto r$, the mass of the circumnuclear pc-scale disk is significant with
respect to the central mass. In fact, the disk mass within $0.7$\,pc, the
distance from the dynamical center to the brightest maser feature, is roughly
$M_d(0.7\mbox{ pc})\sim M(0.7\mbox{ pc})-M(0.4\mbox{ pc})\sim10^6$\,$M_{\sun}$.
The mass of the entire disk traced by the maser emission ($<1.3$ pc) might be as
large as $7\times 10^{6}$\,$M_{\odot}$, but this is uncertain by a factor of a
few. Interestingly, if we extrapolate $M(r)\propto r$ to larger radii, we
estimate a mass of $\sim 4\times10^8$\,$M_{\sun}$ enclosed within $76$\,pc, which
is in surprisingly good agreement with the $7\times10^8\,M_{\sun}$ dynamical mass
inferred from CO observations \citep{Koda2002}.

\subsubsection{Mean Mass Density, Eddington Ratio, and Mass Accretion Rate}
\label{massdensity} The mean mass density corresponding to $2\times
10^6$\,$M_{\odot}$ enclosed within $0.4$\,pc is $10^{6.8}\,M_{\sun}$\,pc$^{-3}$.
The relatively high mean mass density for NGC 3079 is suggestive of a massive
central black hole as opposed to a dense star cluster \citep[e.g.,][]{Maoz1995},
which is consistent with the X-ray observations of the nucleus. The estimated
enclosed mass of $\sim2\times10^6$\,$M_{\odot}$ is in agreement with proposed
correlations between bulge velocity dispersion and black hole mass
\citep[Gebhardt et al. 2000a, 2000b;][]{Ferrarese2000, Ferrarese2001}. If we
adopt $M_{BH}=1.2\times10^8\,M_{\odot}\,[\sigma/(200\mbox{ km s}^{-1})]^{3.75}$
(Gebhardt et al. 2000a, 2000b)\nocite{Gebhardt2000a}\nocite{Gebhardt2000b}, then
the velocity dispersion of the bulge, $40$\,km\,s$^{-1}<\sigma<160$\,km\,s$^{-1}$
\citep*{Shaw1993}, is consistent with a black hole mass of
$10^{5.5-7.7}$\,$M_{\odot}$, although the heavy dust obscuration might bias the
measured dispersion.

To estimate the mass accretion rate as well as the accretion timescale, we first
approximate the luminosity of the nucleus. Since the $2-10$\,keV luminosity is
probably $\sim5\%$ of the AGN bolometric luminosity (e.g., Kuraszkiewicz et al.
2003, Elvis et al. 1994)\nocite{Kuraszkiewicz2003}\nocite{Elvis1994}, we obtain
an AGN bolometric luminosity of $\sim5\times10^{9-10}$\,$L_{\odot}$ from the
X-ray data analysis reported by \cite{Iyomoto2001}. The Eddington luminosity of a
$2\times10^6$\,$M_{\odot}$ object is $7\times 10^{10}$\,$L_{\odot}$ and, assuming
that all of the enclosed mass is concentrated in a supermassive black hole, the
approximate luminosity of the central engine yields an Eddington ratio of
$0.08-0.8$, which is consistent with the $0.01-1$ range obtained for Seyfert 1
galaxies, representative supermassive black hole systems
\citep[e.g.,][]{Padovani1989, Wandel1999}. Assuming a standard accretion
efficiency of $\sim0.1$ (Frank, King, \& Raine 2002; see also Marconi et al.
2004)\nocite{Frank2002}\nocite{Marconi2004}, we obtain a mass accretion rate of
$\dot{M}=0.007\,L_{Bol,10}$\,$M_{\sun}$\,year$^{-1}$, where $L_{Bol,10}$ is the
AGN bolometric luminosity in units of $10^{10}\,L_{\odot}$. The disk mass
computed above yields an average accretion timescale of $t=M_d(r)/\dot{M}\sim
10^{8}\,L_{Bol,10}^{-1}$\,years at a radius of $0.7$\,pc (Table \ref{tab4}).

\begin{table}[!h]
\begin{center}
\caption{Parameters of the central engine and pc-scale disk in NGC 3079.
\label{tab4}} \footnotesize
\begin{tabular}{lr}\tableline\tableline
\emph{Quantity} & \emph{Estimate} \\ \tableline
Central mass ($r<0.4$ pc) & $\sim2\times 10^{6}\,M_{\sun}$ \\
Mean mass density & $\sim10^{6.8}\,M_{\sun}$ pc$^{-3}$ \\
Disk mass ($0.4\mbox{ pc}<r<0.7\mbox{ pc}$) & $\sim 10^{6}\,M_{\sun}$ \\
Disk mass ($0.4\mbox{ pc}<r<1.3\mbox{ pc}$) & $\lesssim7\times 10^{6}\,M_{\sun}$ \\
Disk scale height & $0.05\mbox{ pc}<h<0.5$ pc \\
AGN bolometric luminosity & $\sim5\times 10^{9-10}$ $L_{\odot}$ \\
Eddington ratio & $0.08-0.8$\\
Accretion rate\tablenotemark{b} & $0.007$ $L_{Bol,10}$ $M_{\sun}$
year$^{-1}$ \\
Accretion timescale\tablenotemark{a,b} & $10^{8}\,L^{-1}_{Bol,10}$ years \\
Toomre Q-parameter & $0.01<Q<0.02$ \\
Clump size\tablenotemark{a} & $<0.006$ pc \\
Clump mass\tablenotemark{a} & $<5\times 10^2\,M_{\sun}$ \\
Jeans mass & $0.3-53\,M_{\sun}$ \\
Rotation period\tablenotemark{a} & $4\times 10^3$ years\\
Clump cooling timescale\tablenotemark{a} & $<60$ years \\
Clump free-fall timescale & $10^{2.5-4.0}$ years \\
Clump collision timescale & $10^{4-5}$ years \\
Roche limit\tablenotemark{a} & $n>5.3\times 10^8$ cm$^{-3}$ \\
\tableline
\end{tabular}

\tablenotetext{a}{Computed at the location of the brightest maser feature,
$r=0.7$\,pc.}

\tablenotetext{b}{$L_{Bol,10}$ is the AGN bolometric luminosity in units of
$10^{10}\,L_{\odot}$.}

\end{center}
\end{table}

\subsubsection{Disordered, Thick, and Flared Disk}
\label{disorder} The disk in NGC 3079 is different from the archetypal Keplerian
disk in NGC 4258 in that the velocity structure of the former is much more
disordered. The rotation traced by the maser emission in NGC 3079 is
characterized by relatively large velocity differences across relatively small
areas on the sky. For instance, the velocity dispersion ($\sigma$) as computed
using an averaging radius of $4$\,mas ($0.34$\,pc) ranges from $20$ to
$80$\,km\,s$^{-1}$ (Fig. \ref{sigma}). The maximum velocity difference between
neighboring maser features is $144$\,km\,s$^{-1}$ across a region as small as
$0.1$\,pc. The large velocity dispersion is most likely indicative of macroscopic
random motions among the molecular clumps responsible for the maser emission
rather than turbulence within the clumps (Section \ref{clump}). As already noted
above, the variability in the angular distribution and in the spectrum of the
maser (Figs. \ref{spectrum}, \ref{maser_zoom}) is a direct consequence of these
macroscopic random motions. Although a significant fraction of the orbital
velocity, the computed velocity dispersion is everywhere much smaller than the
escape velocity of a $\sim2\times 10^6\,M_{\sun}$ central object
($v_{escape}>110$\,km\,s$^{-1}$ in the region supporting maser emission) and thus
has little impact on the stability of the underlying rotating structure. Although
we favor the macroscopic random motions of clumps as the origin of the observed
high dispersion, we note the small size of the dominant maser feature
($<5\times10^{16}$\,cm) for which $\sigma\sim14$\,km\,s$^{-1}$, and we speculate
that there may be some regions in which turbulence is significant but does not
interfere with maser action \citep*[e.g.,][]{Wallin1998}.

The computed velocity dispersion can be used to infer the disk thickness. Because
$\sigma$ is a considerable fraction of the orbital velocity, the rotating
structure is most likely geometrically thick \citep*[e.g.,][]{Alves2000,
Wainscoat1989}. Moreover, since $\sigma$ does not decrease with the impact
parameter, the disk scale height ought to increase with the distance from the
center; i.e., the disk is probably flared. In particular, the scale height,
computed using $h\sim r\sigma/\langle v\rangle$ appropriate in the case of
hydrostatic equilibrium, is on the order of $0.05$\,pc at an impact parameter of
$0.5$\,pc and might be as large as $h\sim 0.5$\,pc at a radius of $0.9$\,pc (Fig.
\ref{sigma}). The computed scale height is in reasonable agreement with the
apparent thickness of the disk inferred from the dispersion of the maser spots
about the disk major axis, $h\sim 0.25$\,pc at $r\sim 0.7$\,pc (see Fig.
\ref{maser2}).

\subsection{Clumpy Star-Forming Disk}
\label{clump}
The Toomre-$Q$ parameter characterizes the stability of the accretion disk, and
we can estimate it based on our measurements and an assumption about the
temperature of the maser medium. The $Q$ parameter is given by $Q = \Omega
c_s/(\pi G \Sigma)$, where $\Omega$ is the angular rotation rate, $\Sigma$ is the
surface density, and $c_s$ is the sound speed. If we assume that the mass
distribution in the accretion disk has circular symmetry, then $\Sigma =
(dM/dr)/(2\pi r)$ and we can write $Q = 2 v c_s/(G\,dM/dr)$. Using $M(r)=v^2r/G$
and the fact that the orbital velocity of $v\sim150$\,km\,s$^{-1}$ is
approximately constant with radius (Fig. \ref{sigma}), we estimate a radius
independent value of roughly $dM/dr = v^2/G \sim 3\times 10^{21}$\,g\,cm$^{-2}$.
The sound speed in a neutral medium is given by $c_s=0.04\,T^{1/2}$\,km\,s$^{-1}$
\citep[e.g.,][]{Maoz1998}, and most models of maser emission from water vapor
require the gas temperature to lie in the range $300-1000$\,K
\citep[e.g.,][]{Desch1998}. With these parameters we obtain $Q=0.01-0.02$, where
the range reflects only the uncertainty in temperature. Except for possible
temperature variation, our estimate of $Q$ is independent of radius since both
the orbital velocity and $dM/dr$ are to first order constant with radius. Since
$Q$ is significantly less than unity, the disk appears to be gravitationally
unstable in the region supporting the maser emission. The flat rotation curve,
which implies significant accretion disk mass with respect to the black hole
mass, is fundamentally responsible for the low value of $Q$. The value of $Q$ we
estimate is rather robust if our fundamental assumption that the velocities of
the masers are due to the gravity of the enclosed mass holds true.

From our estimate of $dM/dr$, we parameterize the disk surface density as $\Sigma
\sim 250\,(r/0.7\mbox{ pc})^{-1}$\,g\,cm$^{-2}$, i.e., about $430$ and
$130$\,g\,cm$^{-2}$ at the inner and outer radii of $0.4$ and $1.3$\,pc,
respectively. (The empirically derived power law index of $-1$ is not
unreasonable; for instance, $\Sigma \propto r^{-3/5}$ for a radiatively cooled,
gas pressure-dominated $\alpha$-disk with a constant rate of accretion throughout
the disk.) Using the scale height in Fig. \ref{sigma}, our estimate for disk
surface density, and $\Sigma=2h\langle\rho\rangle$, we compute mean density that
varies from about $0.3\times 10^8$ to $3\times 10^8\,m_H$\,cm$^{-3}$ over the
observed disk. The densities deduced here lie comfortably within the normally
required range of H$_2$ number densities of $10^{7-10}$\,cm$^{-3}$ for maser
emission \citep[e.g.,][]{Desch1998}. Our determination here of the average
density of a maser medium is probably the first accurate determination of this
quality for water in general.

The ultimate fate of an accretion disk characterized by $Q<1$ depends on the
balance between cooling and heating. It has been argued that the disk will
maintain $Q\sim 1$ by rearrangement of surface density to reduce $\Sigma$ or by
heating through turbulence \citep{Lin1987, Hure2000, Gammie2001}. However, the
energy required to maintain $Q\sim 1$ is prohibitive \citep{Goodman2003} and, in
the case of NGC 3079, would require temperatures as high as $10^6$\,K.
Furthermore, cooling timescales might be sufficiently short that a $Q\sim 1$
criterion is unsupportable \citep{Shlosman1989, Monaghan1991}, in which case the
disk would fragment into clumps \citep[e.g.,][]{Gammie2001, Kumar1999}.

We consider the clumpy disk model in detail by computing clump size and mass at a
representative location within the disk, the distance from the dynamical center
to the brightest maser feature, $0.7$\,pc.  The gas number density required for
gravitational collapse in a strong tidal field is given by the Roche limit, $n
> 5.3\times 10^{8} \,(M(r)/2\times 10^6\,M_{\sun})\, (0.7\mbox{ pc}/r)^3$
cm$^{-3}$, which is consistent with the H$_2$ number density required by the
presence of the maser emission. Furthermore, as derived by \cite*{Vollmer2003},
an upper limit on the radius of a self-gravitating clump subject to tidal shear
is $R_{c} < \pi c_s/\sqrt{8}\Omega$, from which we obtain $R_{c} < 0.006$\,pc and
an upper limit on the clump mass of $5\times 10^2\,M_{\sun}$. It is instructive
to compare this upper limit to the Jeans mass, $m_J=(\pi
c_s^2/G)^{3/2}\rho^{-1/2}$, which is $0.3-53\,M_{\sun}$ given the conditions
necessary to support maser emission. Since the Jeans mass and the Roche limit are
consistent with the upper limit on cloud mass and with H$_2$ number density
inferred from the presence of the maser emission, respectively, fragmentation and
star formation are expected to occur within the clumps.

In addition to the Jeans mass analysis, straightforward considerations of
energetics also suggest that individual clumps are susceptible to collapse. The
time in which a clump will radiate its binding energy via black-body emission is
given by $t_c \sim (GM_c^2/R_c)/(4\pi R_c^2\sigma T^4_c)\approx 2.3\times (GM_c
m_H \langle n_{\rm H_2}\rangle)/(3\sigma T_c^4)$ where $M_c$, $R_c$, $T_c$, and
$\langle n_{\rm H_2}\rangle$ are the clump's mass, radius, effective temperature,
and average H$_2$ number density, respectively \citep[e.g.,][]{Goodman2003}. In
our case, $M_c<4\times 10^2\,M_{\sun}$, $T_c>300$\,K, and $\langle n_{\rm
H_2}\rangle \le 10^{10}$\,cm$^{-3}$, which yields $t_c< 60$\,years. However, if
line emission is the dominant cooling mechanism, then the timescale becomes $t_c
\sim (GM_c^2/R_c)/(4\pi R_c^3\Lambda/3)\approx (2.3)^2\times (4\pi G m_H^2
R_c^2\langle n_{\rm H_2}\rangle/3)(\Lambda/\langle n_{\rm H_2}\rangle)^{-1}$,
where $\Lambda/\langle n_{\rm H_2}\rangle$ is the total cooling rate per H$_2$
molecule in ergs\,s$^{-1}$. Using the cooling rates in \cite{Neufeld1995} and
$R_c<0.006$\,pc, the expression above is maximized for $\langle n_{\rm
H_2}\rangle= 10^{10}$\,cm$^{-3}$ and $T_c=300$\,K, which results in
$t_c<40$\,years. If cooling of the gas by cold dust grains dominates, then, using
$R_c<0.006$\,pc, dust temperature of $T_d\approx 200$\,K \citep{Collison1995},
and the cooling rate per unit volume of \cite{Hollenbach1989}, we obtain a
cooling timescale of at most $5$\,years. Thus, the time in which a clump will
radiate all of its binding energy via either black-body radiation, line emission,
or heating of cold dust is much less than the accretion timescale and even the
disk rotation period (Table \ref{tab4}); consequently, clumps collapse on their
free-fall timescale, resulting in star formation. In our case, the free-fall
timescale is $\sqrt{R_c^3/GM}\approx(3\pi G m_H n_{\rm
H_2})^{-0.5}=10^{2.5-4.0}$\,years, depends only on the hydrogen number density
implied by the conditions necessary for maser amplification, and is four orders
of magnitude smaller than the computed accretion timescale.

Star formation can also be triggered by collisions among clumps. W49N, the most
luminous water maser in our Galaxy, provides probably the best example of how
clump-clump interaction can induce O-type star formation \citep*{Serabyn1993}. To
estimate the collision timescale, we must obtain the mean clump separation. Using
the scale height in Fig. \ref{sigma}, we estimate the volume of the entire disk
as traced by the maser emission to be roughly $2$\,pc$^3$. Mass of the entire
maser disk of $\sim7\times10^6$\,$M_{\odot}$ and a $5\times10^2$\,$M_{\odot}$
upper limit on clump mass yield $>1.5\times 10^4$\,clumps in the disk and mean
clump separation of $l_s<0.03$\,pc. Hence, the collision timescale, $\sim
\Omega^{-1} \,(l_s/R_{c})^2$ \citep{Kumar1999}, is $\sim 10^{4-5}$\,years and in
fact depends only on disk size, disk mass, and H$_2$ number density inferred from
the presence of the maser emission. Since the collision timescale is much less
than the accretion timescale, collisions among clumps might also be a significant
process in triggering star formation and overall evolution of the disk.

In summary, since $Q\ll1$, the pc-scale circumnuclear disk in NGC 3079 is most
likely gravitationally unstable and therefore clumpy. Evidence for a
self-gravitating disk of mass comparable to the central mass has also been found
in NGC 1068 \citep[$M_d\approx M_{BH} = 8.0\times
10^6\,M_{\odot}$;][]{Lodato2003}, although NGC 3079 provides a more extreme
example of the phenomenon ($M_d\sim 7\times 10^6\,M_{\odot}$ vs. $M_{BH} \sim
2\times 10^6\,M_{\odot}$). Furthermore, the Jeans mass, the Roche limit, as well
as cooling and collision timescales are suggestive of star formation within the
clumps through either clump collapse or clump collisions. Once a clump collapses,
star formation proceeds via accretion of gas from the pc-scale accretion disk
onto the Hill sphere of the stellar object, as considered in detail by
\cite{Milosavljevic2004}. The star formation efficiency of giant molecular clouds
($M=10^{1-3}\,M_{\sun}$) and dark molecular cores ($M=1 \,M_{\sun}$) ranges from
$0.1$\,\% to $5$\,\% while the star formation timescale and the lifetime of hot
massive stars are on the order of $10^6$\,years \citep[e.g.,][]{Blitz1980,
Larson1982, Wilking1983, Myers1985, Silk1985, Silk1987}. Thus, the long accretion
timescale and the large potential clump mass computed above are consistent with
the relatively slow and inefficient process of star formation (Table 4). If the
star formation were to exhaust the available gas (which, in addition to the star
formation rate, also depends on the presence of any gas feeding mechanisms), then
the massive accretion disk in NGC 3079 would transform into a stellar disk, a
transition that has been proposed to explain the presence of young stars in the
immediate vicinity of the black hole in the Galactic center \citep{Levin2003,
Milosavljevic2004, Nayakshin2004}.

\subsection{Pumping of Maser Emission by the Central Engine}
Since the nucleus of NGC\,3079 contains a compact hard X-ray source, irradiation
of molecular gas is a plausible means of exciting maser emission
\citep*[e.g.,][]{Neufeld1994}. In maser sources such as NGC\,4258 and Circinus,
irradiation over a wide range of radii is achievable because the disks are warped
(Herrnstein, Greenhill, \& Moran 1996; Greenhill et al.
2003)\nocite{Greenhill2003}\nocite{Herrnstein1996}. In NGC\,3079, the disk in
which the masers lie does not appear to be warped, in which case maser excitation
over the observed range of radii ($0.4-1.3$\,pc) depends on penetration of X-rays
in the disk plane. However, because maser action requires high densities, the
disk must be inhomogeneous, which is consistent with our earlier stability
arguments. For a minimum H$_2$ density of $10^7$\,cm$^{-3}$, a column within the
disk becomes Compton thick for lengths greater than $\sim0.02$\,pc. However, for
a line-of-sight filling factor of $\sim2\%$, the column could pass X-rays to a
radius of $\sim 1.3$\,pc.

For an irradiated slab of gas, \cite{Collison1995} estimate a maser emission rate
of $\sim4000$\,$L_{\odot}$ per pc$^{2}$ of surface area beamed in the plane of
the slab. For a model spherical gas clump of radius $\sim0.005$\,pc that is
irradiated on one side, the integrated output is on the order of
$0.3$\,$L_{\odot}$, which is much less than the isotropic luminosity inferred
from the strength of individual maser features (e.g., $131$\,$L_{\odot}$;
Fig.\ref{spectrum}). We suggest instead that the emission from the observed
masers is narrowly beamed along our line of sight by the overlap of clumps with
similar Doppler velocities, whereby the integrated output of each is directed
toward us \citep[e.g.,][]{Deguchi1989, Kartje1999}. To support the peak observed
isotropic luminosity, we require two clumps, each $0.005$\,pc in size, separated
by on the order of $0.1$\,pc, which is reasonable in light of the $\sim1.3$\,pc
radius of the disk. We note that amplification of background continuum emission
by individual clumps can also generate strongly forward beamed maser emission,
but we do not detect continuum emission in the vicinity of the disk above
$0.84$\,mJy ($6\sigma$) at $22$\,GHz. On the other hand, the two easternmost and
westernmost maser features do lie in close vicinity to continuum components B and
C, and we speculate that maser emission observed away from the disk, where the
density of clumps is probably greatly reduced, may be the result of continuum
amplification.

Under the assumption that clumps are optically thick, irradiation of individual
clumps by a central source creates a dissociation region on the inward facing
side, which results in anisotropic emission of microwave photons preferentially
along the length of the dissociation region. This pattern may be used to explain
why we observe maser emission in the vicinity of the disk midline but not close
to the line of sight to the central engine. Clumps on the front and back sides of
the disk beam maser photons along axes that are substantially inclined to our
line of sight. Furthermore, we speculate that the observed variability of maser
spectrum \citep{Nakai1995, Baan1996, Hagiwara2002} and the observed variation in
angular structure (Fig. \ref{maser_zoom}) may be the result of pairs of clumps
moving in and out of alignment along the line of sight. The presence of the maser
feature at $\sim 956$ km s$^{-1}$ in the spectrum of NGC 3079 since the discovery
of the maser \citep{Henkel1984, Haschick1985, Baan1996} and the
$\sigma\sim30$\,km\,s$^{-1}$ dispersion due to bulk random motion
(Fig.\ref{sigma}) place a lower limit on the clump size of $\sim0.001$ pc,
consistent with the upper limit (Table \ref{tab4}). Year-to-year variation in the
flux density of some maser features might be due to incomplete overlaps on the
sky of the X-ray dissociation regions.

\subsection{Outflow}
\label{outflow}
\subsubsection{Components E and F}
Component E, identified by \cite{Kondratko2000}, and component F, identified by
\cite{Middelberg2003}, might be remnants of jet emission along former axes of a
wobbling jet, currently traced by components A, B, and C. The apparent decrease
in the break frequency of components E and F with distance from the central
engine \citep{Middelberg2003} and the shift of the spectrum of component E toward
lower frequencies with time (Section \ref{continuum}) are indirectly supportive
of the hypothesis that both components are rapidly cooling and adiabatically
expanding remnants of jet emission. Although the structure of the continuum is
difficult to understand in detail, we speculate that components A and B are
regions where the jet, in its current orientation, interacts with a dense ambient
medium, an inference supported by their GPS-like spectra (Section
\ref{continuum}).

Alternatively, considering the location of E (Fig. \ref{model}) and of F
\citep{Middelberg2003} within the opening angle of the kpc-scale superbubble,
these two components might provide evidence for a pc-scale wide-angle outflow,
most likely an inward extension of the kpc-scale superbubble. In this
interpretation, the two VLBI components could be moving knots or shocks as the
ionized wide-angle wind interacts with the dense ambient medium along the edges
of the flow as seen in projection. If this hypothesis is true, then new continuum
components are expected to appear on both sides of the central engine but within
the opening angle of the superbubble. However, it is currently not possible to
distinguish between the two models proposed above, and more sensitive images and
proper motion studies of the continuum components extending over a longer time
frame are necessary. Given the possible source lifetimes and transverse speeds
(Section \ref{continuum}), time baselines on the order of a decade are necessary.

On the other hand, it is unlikely that E and F are radio supernovae associated
with the nuclear starburst. The spectral index of E, $\alpha< -2.1$ between 8 and
15 GHz (Fig. \ref{spectral_index}), is much steeper than what has been observed
in radio supernovae ($\alpha \gtrsim -0.9$; Weiler et al. 1986; Weiler, Panagia,
\& Sramek 1990; Allen \& Kronberg 1998; McDonald et al. 2001, 2002; Bartel et al.
2002)\nocite{Weiler1986}\nocite{Weiler1990}\nocite{Allen1998}
\nocite{McDonald2001}\nocite{McDonald2002}\nocite{Bartel2002}. The extremely
steep spectrum of F, $\alpha\sim-6.1$ between 1.7 and 2.3\,GHz, is also not
compatible with a radio supernova. Furthermore, the radio supernovae in the
prominent starburst galaxy M82 scaled to the distance of NGC 3079 would have flux
densities ($\lesssim 0.1$\,mJy) below our detection limits at 5 and 15\,GHz
\citep{McDonald2002}.

\subsubsection{Wide-Angle Outflow on Parsec Scales}
\label{woa} The existence of a wide-angle outflow in NGC 3079 was hypothesized in
previous studies of the nucleus. Based on VLA data, \cite{Duric1988} first
suggested that a wide-angle wind is responsible for a bipolar ``figure eight"
structure along the minor axis of the galaxy in radio continuum. A wide-angle
outflow on pc-scales has been suggested to explain a blue-shifted OH absorption
component and two weak OH emission features detected from the nucleus
\citep{Baan1995, Hagiwara2004}. By considering the energetics and morphology of
the kpc-scale superbubble observed in H$\alpha$ with the Hubble Space Telescope
(HST), \cite{Cecil2001} argued that the bubble is inflated by a wide-angle
outflow rather than a precessing jet. Wide-angle outflows seem to be quite common
among Seyfert galaxies \citep{Colbert1996a,Colbert1996b,Colbert1998}, and one has
been imaged on pc-scales in the Circinus galaxy, the nearest Seyfert 2 nucleus
(Greenhill et al. 2003)\nocite{Greenhill2003}. Guided by this result, by the
presence of molecular gas as dense as the disk gas at high latitudes above the
disk (i.e., the two easternmost and two westernmost maser features in Fig.
\ref{maser2}), and by the existence of a kpc-scale superwind, we suggest a
wide-angle outflow on pc-scales in NGC 3079.

It has been suggested that winds are driven by photoionized evaporation of matter
from an inner surface of a torus \citep{KrolikKriss2001}, by radiation and gas
pressure acting on an accretion disk \citep{Murray1995}, or by the
magneto-centrifugal uplift of gas and dust from an accretion disk
\citep*[][Kartje et al. 1999]{Emmering1992, Konigl1994}. In addition, ram
pressure of a wide-angle outflow can entrain clumps at the surface of an
accretion disk. If the wind is sufficiently dense to effectively shield the
clumps from the central engine and the clumps can be confined, then they can
potentially rise to high latitudes above the rotating structure \citep[see Fig. 7
in][]{Kartje1999} while still maintaining, to some degree, the rotational
velocity imprinted by the parent disk. In light of this, the fact that the
line-of-sight velocities of the two easternmost and the two westernmost features
(Fig. \ref{maser2}) reflect the velocity of the most proximate side of the disk
can be explained if these clumps were uplifted from the disk surface and carried
to high latitudes by a dense wide-angle outflow. We thus consider the four maser
features located significantly out of the plane of the disk as indirect evidence
for a pc-scale wide-angle outflow, likely an inward extension of the kpc-scale
superbubble. The fact that some clouds attain large latitudes above the disk
surface while others remain within or in a close proximity to the thick disk
might be due to a mixture of heavy and light clumps within the disk. We note
that, due to beaming effects of maser emission, the actual number of the clumps
uplifted might be much greater than the four clumps that are observed to be
associated with the wind.

Although the origin of the putative pc-scale wide-angle outflow is unclear, we
suggest that it is unlikely to be driven by star formation in the pc-scale
accretion disk. The mechanical luminosity required to blow out a superbubble from
the galactic disk in NGC 3079 is $\sim3\times10^{41}$\,ergs\,s$^{-1}$
\citep{Veilleux1994}. Following \cite{Strickland2004b}, such a luminosity would
require $\sim 10^{4}\,(t/\mbox{Myr})$\,supernovae corresponding to a cluster of
mass $5\times 10^{5}\,(t/\mbox{Myr})$\,$M_{\odot}$, assuming a thermalization
efficiency of $\sim1$, an energy release of $10^{51}$\,ergs\,per\,supernova, and
a starburst lifetime $t$ equal to the dynamical age of the superbubble,
$\sim10^6$\,years \citep{Veilleux1994}. The predicted cluster mass is an order of
magnitude greater than the $21,000-79,000$\,$M_{\odot}$ mass of one of the most
massive star clusters known, R136 in 30 Doradus, with a half-mass radius of
$0.5$\,pc \citep[Campbell et al. 1992; Brandl et al.
1996;][]{Massey1998}\nocite{Brandl1996}\nocite{Campbell1992}. Such massive
clusters are thus extremely rare, constituting only a fraction of $\sim5\times
10^{-5}$ of all star clusters \citep{Strickland2004b}. More reasonable values of
the starburst lifetime ($\sim40$\,Myr) would require an even more extreme cluster
mass and number of supernovae, effectively excluding a pc-scale cluster as a
progenitor of the kpc-scale superbubble. Supernovae localized in multiple
starburst regions and distributed over larger scales (e.g., kpc) might still
contribute significantly to the wide-angle outflow, although the degree of this
contribution is uncertain \citep{Strickland2004a, Strickland2004b}.

\begin{figure*}[!h]
\centerline{\includegraphics[width=7.2in]{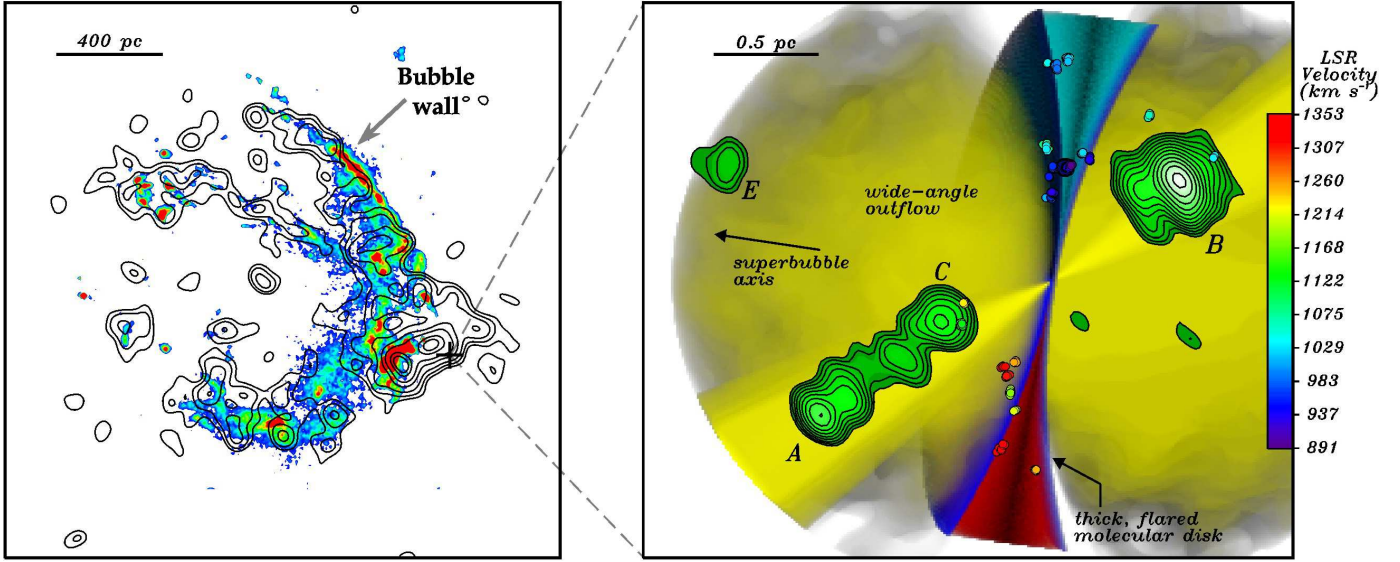}} \figcaption{{\it Left:} [NII]
+ H$\alpha$ image of the superbubble (filled color contours) obtained with HST
\citep{Cecil2001} and the soft X-ray image of the same (black contours) obtained
with Chandra. The X-ray image has been convolved with a Gaussian of $0.64\arcsec$
FWHM. The cross indicates both the VLBI-derived position of the maser source with
respect to the X-ray emission and the $0.6\arcsec$ uncertainty in the Chandra
absolute astrometry (Chandra X-Ray Center Web site:
http://cxc.harvard.edu/cal/ASPECT/celmon/). Note that the position of a nearby
soft X-ray emission peak, $\alpha_{2000}=10^{h}01^{m}57\fs91\pm0.06$,
$\delta_{2000}=55\degr40\arcmin47\farcs7\pm 0.6$, is offset from the dynamical
center of the maser disk by $\sim1\arcsec$. The alignment of the two images has
been accomplished by computing their two-dimensional cross-correlation. Although
the uncertainty in the HST absolute positions might be as large as a few
arcseconds, Monte Carlo simulations of the two-dimensional correlation, performed
by repeatedly adding Gaussian or Poisson noise to the original images, indicate
that the two images are registered to within $0.16\arcsec$. {\it Right:} Proposed
model for the nuclear region of NGC 3079, a VLBI image of 8\,GHz continuum (see
Fig.\ref{8GHz}), and a map of 22\,GHz maser emission (shown as circles
color-coded by Doppler velocity). We propose a thick and flared disk (shown in a
cross-section) that is slightly tilted with the west side being closer and that
is aligned with both the kpc-scale molecular disk and the axis of the kpc-scale
superbubble. The color-coding indicates the Doppler shift of the disk material.
The water maser emission traces the disk and may arise from X-ray irradiated
clumps within the disk, which is unstable to fragmentation. An off-axis jet
traced by the continuum components A, B, and C probably coexists with a
wide-angle outflow (yellow), most likely an inward extension of the kpc-scale
superbubble. The outflow may result in an uplift of clumps from the disk surface,
which would explain the high-latitude maser features and their association in
velocity with the most proximate side of the disk.\label{model}}
        \hrulefill\
\end{figure*}

One consequence of the broad outflow that we hypothesize --- or the canted jet
marked by continuum components A, B, and C --- may be the uniform weakness of
maser emission south of the dynamical center. Since AGN outflows are ionized,
differential attenuation of the maser emission is readily achieved for free-free
absorption with a small difference in electron density and corresponding emission
measure. If the disk is tilted slightly from edge-on such that the lines of sight
to the maser emission traverse outside of the disk (Fig.\ref{model}), the
observed two order of magnitude flux density difference between blue- and
red-shifted features may be explained by a factor of at most $\sim2$ difference
in local electron number density (provided that $\tau>1$ in the direction of the
blue-shifted maser emission). In the case of a canted jet, an enhancement in
density may be readily achieved given that the jet is inclined toward the
apparently weaker masers \citep[see also][]{Trotter1998}, but a density
inhomogeneity in the wide-angle outflow might also be the cause of the
differential attenuation. However, in either case, it is interesting to note that
the balance of blue- and red-shifted emission can be readily governed by factors
extrinsic to the maser emitting regions in the case of both NGC\,3079 and
NGC\,4258 \citep{Herrnstein1996}. This may be generalizable to other maser
sources and contrasts with the models of \cite{Maoz1995b} and \cite{Maoz1998}.

\section{Summary}

Water maser emission in the active nucleus of NGC\,3079 has recently been
recognized to cover a $\sim 450$\,km\,s$^{-1}$ range centered on the systemic
velocity of the galaxy.  We have mapped for the first time maser emission over
this entire velocity interval. We have also imaged non-thermal continuum emission
between 5 and 22\,GHz that arises in close proximity to the molecular gas
underlying the masers. Based on the analysis of the spectral-line and continuum
maps, we conclude the following:

\begin{itemize}
\item[1.] The largely north-south distribution of maser emission, aligned with a
known kpc-scale molecular disk, and the segregation of blue- and red-shifted
maser emission on the sky strongly support the model of \cite{Trotter1998} in
which the masers trace a nearly edge-on molecular disk about $1$\,pc in radius.

\item[2.] The dynamical mass enclosed within a $0.4$\,pc radius is
$2\times10^6\,M_\odot$, and for a $40$ to $160$\,km\,s$^{-1}$ stellar bulge
velocity dispersion is consistent with correlations between central mass and
dispersion reported for broad samples of galaxies \citep[Gebhardt et al. 2000a,
2000b;][]{Ferrarese2000, Ferrarese2001}. The ratio of bolometric to Eddington
luminosity is $0.08-0.8$, consistent with accretion efficiencies of $0.01-1$ for
Seyfert 1 systems.

\item[3.] The disk rotation curve is relatively flat, which is consistent with a
mass of $\lesssim7\times10^6\,M_{\odot}$ between radii of $0.4$ and $1.3$\,pc.

\item[4.] The angular distribution of maser emission is not as well ordered in
NGC\,3079 as in other ``maser galaxies'' (e.g., NGC4258). The velocities of
adjacent clumps of maser emission can differ by tens of km\,s$^{-1}$ on scales of
$\sim 0.1$\,pc. As a result, we suggest that the disk is relatively thick ($h/r
\sim 0.1$ to $0.5$) and may be flared.

\item[5.] Based on stability, cooling, and timescale arguments, we argue that the
disk is self-gravitating, clumpy, and appears to meet the necessary conditions
for star-formation. The maser emission most likely occurs as a result of clumps
irradiated by X-rays from the central engine and overlapping on the sky, yielding
long gain paths and narrow beam angles.

\item[6.] We report detection of a very steep spectrum synchrotron component (E)
that is not collinear with the previously claimed compact jet. The spectrum is
consistent with an aging electron energy distribution. The observation of a
``relic'' favors the hypothesis that the jet in NGC\,3079 has changed direction.
In addition to the jet, we observe molecular gas that is dense enough to support
maser emission at high latitudes above the disk. From this, we infer that the jet
coexists with a wide-angle outflow originating at parsec or smaller scales.

\end{itemize}

The proposed model can be tested through further VLBI study of the NGC\,3079 AGN.
More sensitive images and proper motion studies of the continuum components, with
a longer time baseline, may help to determine the nature of components E and F.
If they are shocks in a wide-angle outflow rather than remnants of jet emission,
then new continuum components are expected to appear within the arc subtended by
the superbubble. The presence of star-formation in the inner parsec can be
confirmed with an ultrasensitive search for M82-like radio supernova remnants. A
proper motion and monitoring study of the maser emission would be challenging
(i.e., motions $<2$\,$\mu$as\,yr$^{-1}$) but could be used to corroborate the
proposed geometry for the inner parsec, although it might be difficult to
disentangle motions due to kinematics from apparent motions due to local effects,
such as clump alignments. Hosting a central engine, a jet, a thick and
self-gravitating accretion disk, and possibly a wide-angle outflow and
star-formation, the nucleus of NGC 3079 constitutes a nearby laboratory for
diverse astrophysical phenomena and is a strong candidate for the study of the
starburst-AGN connection and of the complex interactions between gas, stars, and
supermassive black holes.

We would like to thank Gerald Cecil for providing HST images in digital form and
Craig Heinke for help in processing the Chandra image. We thank the anonymous
referee for useful comments.

\bibliography{ms}

\end{document}